\documentclass[11pt]{article}
\usepackage{fullpage}
\usepackage{amsmath}
\usepackage{amssymb}
\usepackage{amsthm}
\usepackage{amsfonts}
\usepackage{bm}
\usepackage{enumitem}
\usepackage[final]{graphicx}
\usepackage{caption}
\usepackage{float}
\usepackage{epstopdf}
\usepackage{subcaption}
\usepackage{multirow}
\usepackage{multicol}
\usepackage{booktabs}
\usepackage{algorithm}
\usepackage{algpseudocode}
\usepackage{url}
\usepackage{cite}
\usepackage{textcomp}
\usepackage{subcaption}
\usepackage[font=small,labelfont=bf]{caption}
\usepackage{multirow}
\usepackage{rotating}
\usepackage{microtype}
\usepackage{authblk}
\usepackage{url}
\usepackage{setspace}
\newtheorem{remark}{Remark}

\author[*]{Siddharth Patel}
\author[*$\dag$]{Ram Rajagopal}
\affil[*]{Department of Civil and Environmental Engineering, Stanford University}
\affil[$\dag$]{Department of Electrical Engineering (by Courtesy), Stanford University}

\date{}

\title{Peer to Peer Sharing of Distributed Energy Resources}
\begin{document}

\maketitle

\begin{abstract}
As the penetration of distributed energy resources in the residential sector increases, the scope for sharing arrangements expands.
We model a peer-to-peer rental market for rooftop solar and energy storage in the residential sector, with households seeking to minimize their electricity costs.
For varying adoption levels, we characterize the market rental price, quantity, and participation rate.
We find that up to 15\% adoption, the peer-to-peer market generates a surplus comparable to that attainable though a centralized sharing model.
The peer-to-peer market can incentivize an increase in total adoption in the long run.
We find that direct subsidies would be a cheaper way to increase adoption if enabling the peer-to-peer market increases distribution grid costs by more than a few percent.
This cost increase would be related to how locally the peer-to-peer market can match renters and owners.
We compute metrics of this localness and find that the market clears quite locally for a wide range of adoption rates.
\end{abstract}

Distributed energy resources (DERs) such as rooftop photovoltaic (PV) systems and energy storage devices promise to fundamentally alter the electric grid by decarbonizing and decentralizing energy services \cite{casa2013,akorede2010,jain2017} and transforming end users from passive loads into active participants \cite{parag2016,bayram2017,agnew2015}.
Declining prices for DERs, combined with rising environmental awareness and high-profile newly commercially available products such as Tesla's Powerwall and Solar Roof, make it likely that adoption of these technologies in the household sector will continue to grow \cite{eia2015,eia2016,deutsche2016,nykvist2015,bronski2015}.
Net-zero policy goals and directives in Europe and California will add to this momentum \cite{CEC2015,CECRelease2018,sartori2012,annunziata2013}.

The value of DERs to households varies based on consumption patterns \cite{patel2017}, so the adoption of DERs has been and will continue to be an uneven process.
This unevenness raises the possibility of sharing arrangements.
Proposals like community solar focus on joint ownership of DERs by a group of households who as individuals may not be able to afford fixed costs or other limitations of small scale \cite{funkhouser2015,feldman2015}.
However, some households, like those who rent apartments or homes, may never choose to own DERs but may still desire to obtain some of the benefits they provide.
Siting considerations like roof angle or size may also limit DER ownership.
On the other hand, owners of DERs may not be able to fully utilize the capacity of their technology or may be willing to share their capacity with others.
The expansion of digital platforms for peer-to-peer markets has enabled more people to share and trade resources such as housing, vehicles, and labor on a smaller scale than previously possible, and this model can extend to DERs \cite{cardwell2017}.
Inspired by Horton and Zeckhauser \cite{horton2016}, we develop a methodology for evaluating a peer-to-peer market for sharing DERs, in which owners of DERs can rent out some of their capacity to non-owners.

We draw upon a large set of hourly electricity consumption data for households in Northern California.
Households may derive many benefits from DERs, like having backup power during outages or reducing their carbon footprint.
In our study, we focus on modeling the value of DERs as the bill savings that they can achieve.
We estimate demand and supply functions for the peer-to-peer market at different levels of DER adoption, and we characterize the resulting equilibria in terms of rental price, household participation rates, and surpluses.
We compare the total market surplus across all households with the surplus achievable in a more centralized model for sharing as formulated in \cite{patel2017}.

The existence of a peer-to-peer market may lead to increased DER adoption.
DER equipment vendors stand to gain from increased adoption, whereas retail electric utilities stand to lose.
We model this opposition between vendors supporting the emergence of the peer-to-peer market and utilities trying to block it, and we provide conditions under which either party prevails.

Direct subsidies are an alternative means to increase DER adoption that may involve less complexity and cost for distribution system operators.
We compute the direct subsidy that would increase DER adoption by the same amount that the peer-to-peer market does.
We compare that figure to potential increases in distribution grid costs that enabling a peer-to-peer market might entail.
These increases are related to the geographical localness of sharing transactions on the peer-to-peer market, so we develop and compute measures of this localness.
This analysis allows policymakers to evaluate the trade-off between the peer-to-peer market and direct subsidies.

\section{Model}

\subsection{Household savings}

We consider a population of households.
One type of DER asset is available, and households decide whether or not to own one of this asset.
The asset is a combination of a PV system sized to be net-zero and a storage device whose capacity is scaled linearly to the PV system size.
Because energy consumption varies among households, different households will have different sized DER assets.
Let $\overline{y}_i$ denote the net-zero PV system size for household $i$.
The storage capacity for household $i$ is then $\alpha\overline{y}_i$, where $\alpha$ is a scaling coefficient that is the same for all households.
Similarly, the maximum charging rate for the storage device is $\overline{u}\overline{y}_i$, and the maximum discharging rate is $\underline{u}\overline{y}_i$, where $\overline{u}$ and $\underline{u}$ are scaling coefficients that are the same for all households.
Thus a single parameter $\overline{y}_i$ defines the net-zero DER asset size for a given household.

The asset enables the household to save money on its electricity cost. The household has an inflexible end-use electrical load $\textbf{L}_i=[\textbf{l}_i^{(1)}\cdots\textbf{l}_i^{(j)}\cdots\textbf{l}_i^{(D)}]$, where $i$ indexes the household and $j$ indexes the day out of $D$ total days.

A retail utility provides electrical service to the household.
The household faces a tariff under which it purchases electricity at prices $\textbf{Q}_i$ and sells surplus electricity back to the retailer at $\textbf{R}_i$, where $\textbf{q}_i^{(j)}$ and $\textbf{r}_i^{(j)}$ are prices on the $j$th day.
The tariff is exogenous --- it is affected neither by the number of households who own the asset nor by what they do with it.

A rental market makes it possible for non-owners to rent asset capacity from owners.
Non-owners can rent at most their net-zero system size, and owners can rent out at most the full use of their asset.
The rental contract lasts the entire period under consideration.

Suppose that household $i$ has capacity $y_i$ of the asset for its own use --- in other words, it is either a non-owner that has rented $y_i$ on the market, or an owner that has rented out $\overline{y_i}-y_i$ on the market and reserved $y_i$ for its own use. Let $\textbf{v}_i^{(j)}$ be the solar irradiance experienced by household $i$ on day $j$.
The net load for the household on this day is then $\textbf{l}_i^{(j)}-\eta_I\textbf{v}_i^{(j)}y_i$, where $\eta_I$ is the efficiency of the inverter connecting the PV to the house. 
We follow \cite{patel2017} in setting up a daily optimization problem for the household in which it minimizes its daily cost of electricity when it selects $\textbf{u}^{(j)}$, the charging and discharging actions of the storage device:

\begin{subequations}
\label{eqn:Optimization}
\begin{align}
\underset{\textbf{u}^{(j)},\textbf{x},\textbf{g} \; \in \; \mathbb{R}^{24}}{\text{minimize}} \; & [\textbf{g}]_+^T\textbf{q}_i^{(j)} + [\textbf{g}]_-^T\textbf{r}_i^{(j)} \\
\mbox{subject to} \; & \textbf{g}=\textbf{l}_i^{(j)}-\eta_I\textbf{v}_i^{(j)}y_i+\frac{1}{\eta_C\eta_I}[\textbf{u}^{(j)}]_++(\eta_D\eta_I)[\textbf{u}^{(j)}]_- \\
& -\underline{u}y_i \leq \textbf{u}^{(j)} \leq \overline{u}y_i \\
& 0 \leq \textbf{x} \leq \alpha y_i \\
& \textbf{x}_0 = x_0 \\
& \textbf{x}_h = \eta_S\textbf{x}_{h-1}+\textbf{u}_h^{(j)},\; \forall h \in \lbrace1,\ldots,24\rbrace.
\end{align}
\end{subequations}

The available storage device capacity, its maximum charging and discharging rates, and the available PV energy are now affected by how much of the asset the household has use of.
The charging and discharging efficiencies $\eta_C$ and $\eta_D$ and the self-discharge efficiency $\eta_S$ are not affected by $y_i$.
The storage state of charge is denoted by $\textbf{x}$, with $\textbf{x}_h$ denoting the state at the end of hour $h$.
The initial state of charge is $x_0$, and the load on the grid is $\textbf{g}$.
The $[\cdot]_+$ operator is element-wise application of $\textrm{max}(\cdot,0)$, and $[\cdot]_-$ is element-wise $\textrm{min}(\cdot,0)$.

Let the value of the objective function at the optimum be $c_i^{(j)}(y_i)$, and $b_i(y_i)=\sum_{j=1}^{D}c_i^{(j)}$.
In words, $b_i(y_i)$ is household $i$'s total electricity bill when operating fraction $y_i$ of the DER asset optimally.
If $y_i=0$, its electricity cost for the period under consideration is $b_i(0)=b_{BL,i}=\textbf{L}_i^T\textbf{Q}_i$.
Define the household savings function $f_i(y_i)=b_{BL,i}-b_i(y_i)$.
This function returns how much the household will save on its electricity bill compared to the baseline if it has use of $y_i$ of the asset; $f(0)=0$, and $f(\overline{y}_i)$ is the most that the household can save. 

\begin{remark}
The savings function is non-decreasing. If on any day the sale price is positive when the irradiance is positive, then the savings function is strictly increasing.
\end{remark}
\begin{proof}
Consider any day $j$ for household $i$.
Assuming $\textbf{q}_i^{(j)}\geq\textbf{r}_i^{(j)}$, which precludes instantaneous and unlimited arbitrage, Optimization \ref{eqn:Optimization} is feasible and bounded.
Suppose $\tilde{\textbf{u}}^{(j)}$, $\tilde{\textbf{x}}$, and $\tilde{\textbf{g}}$ are optimal for $y_i=\tilde{y}_i$.
The optimal objective value for this value of $y_i$ is then $c_i^{(j)}(\tilde{y}_i)=[\tilde{\textbf{g}}]_+^T\textbf{q}_i^{(j)} + [\tilde{\textbf{g}}]_-^T\textbf{r}_i^{(j)}$.
Now consider the optimization with $y_i=\tilde{y}_i+\delta$, where $\delta>0$.
Both $\tilde{\textbf{x}}$ and $\tilde{\textbf{u}}^{(j)}$ still satisfy constraints (c) through (f), and $\hat{g}=\tilde{g}-\delta\eta_I\textbf{v}_i^{(j)}$ satisfies constraint (b).
The irradiance is nonnegative, so $\hat{g}\leq\tilde{g}$.
The prices are nonnegative, so the objective value at this feasible $\hat{\textbf{g}}$ is less than or equal to $c_i^{(j)}(\tilde{y}_i)$, which implies that the optimal objective value $c_i^{(j)}(\tilde{y}_i+\delta)\leq c_i^{(j)}(\tilde{y}_i)$.
This is true for all $j$, so $b_i(\tilde{y}_i+\delta)\leq b_i(\tilde{y}_i)$, and $f_i(\tilde{y}_i+\delta)\geq f_i(\tilde{y}_i)$.

Now suppose for some hour $\hat{h}$ of day $\hat{k}$, $\textbf{v}_{i,\hat{h}}^{(\hat{k})}>0$ and $\textbf{r}_{i,\hat{h}}^{(\hat{k})}>0$.
Under the assumption of no instantaneous, unlimited arbitrage, $\textbf{q}_{i,\hat{h}}^{(\hat{k})}\geq\textbf{r}_{i,\hat{h}}^{(\hat{k})}$.
The objective value for $y_i=\tilde{y}_i+\delta$ at the feasible $\hat{\textbf{g}}$ is now less than or equal to $c_i^{(\hat{k})}(\tilde{y}_i)-\delta\eta_I\textbf{v}_{i,\hat{h}}^{(\hat{k})}\textbf{r}_{i,\hat{h}}^{(\hat{k})}$, so the optimal objective value $c_i^{(\hat{k})}(\tilde{y}_i+\delta) < c_i^{(\hat{k})}(\tilde{y}_i)$.
This implies that $f_i(\tilde{y}_i+\delta)>f_i(\tilde{y}_i)$.
\end{proof}

\begin{remark}
The savings function is concave.
\end{remark}
\begin{proof}
Consider any day $j$ for household $i$. Under the assumption of no instantaneous, unlimited arbitrage, Optimization \ref{eqn:Optimization} is feasible and bounded.
Let $\bm{\theta}$ be a single decision variable vector that is the concatenation of $[\textbf{u}^{(j)}]_+$, $[\textbf{u}^{(j)}]_-$, $\textbf{x}$, $[\textbf{g}]_+$, and $[\textbf{g}]_-$.
Optimization \ref{eqn:Optimization} can be rewritten as:
\begin{subequations}
\label{eqn:general_LP}
\begin{align}
\underset{\bm{\theta}\; \in \; \mathbb{R}^{120}}{\text{minimize}} \; & \bm{\pi}^T\bm{\theta} \\
\mbox{subject to} \; & A\bm{\theta}+B=Cy_i+D\\
&E\bm{\theta}+F\leq Gy_i+H,
\end{align}
\end{subequations}
for suitable choices of matrices $A$ and $E$ and vectors $\bm{\pi}$, $B$, $C$, $D$, $F$, $G$, and $H$.
Consider $y_i=\tilde{y}_i$ with an optimal solution $\tilde{\bm{\theta}}$ and optimal objective value $c_i^{(j)}(\tilde{y}_i)=\bm{\pi}^T\tilde{\bm{\theta}}$, and a similar tuple of $\hat{y}_i$, $\hat{\bm{\theta}}$, and $c_i^{(j)}(\hat{y}_i)=\bm{\pi}^T\hat{\bm{\theta}}$.
Now consider any convex combination $y_i'=\alpha \tilde{y}_i+(1-\alpha)\hat{y}_i$, with $\alpha \in [0,1]$.
One can show that $\bm{\theta}'=\alpha \tilde{\bm{\theta}}+(1-\alpha)\hat{\bm{\theta}}$ is feasible for Optimization \ref{eqn:general_LP} with $y_i=y_i'$.
The objective value $\bm{\pi}^T\bm{\theta}'=\alpha \bm{\pi}^T\tilde{\bm{\theta}}+(1-\alpha)\bm{\pi}^T\hat{\bm{\theta}}$.
The optimal objective value at $y_i=y_i'$ must be less than or equal to $\bm{\pi}^T\bm{\theta}'$, so $c_i^{(j)}(y_i')\leq \alpha c_i^{(j)}(\tilde{y}_i)+(1-\alpha) c_i^{(j)}(\hat{y}_i)$.
Therefore $c_i^{(j)}(y_i)$ is convex, and $f_i(y_i)$ is concave.
\end{proof}

\subsection{Household demand and supply}

In our model, households make decisions on the rental market based solely on maximizing their savings on electricity.
We ignore other benefits that DERs may bring to a household, like backup power during outages, the enjoyment of actively participating on the grid, or the satisfaction of using renewably generated electricity \cite{roe2001,zarnikau2003,ma2015}.
Given an exogenous per-unit rental price $r$, a non-owner household solves the following optimization:
\begin{equation}
\underset{y_i\;\in\;[0,\overline{y}_i]}{\text{maximize}} \; f_i(y_i)-ry_i,
\label{eqn:Non-owner}
\end{equation}
whereas an owner household solves:
\begin{equation}
\underset{y_i\;\in\;[0,\overline{y}_i]}{\text{maximize}} \;f_i(y_i)+r(\overline{y}_i-y_i).
\label{eqn:Owner}
\end{equation}

Let $y_i^\star(r)$ be household $i$'s optimal choice of $y_i$ given rental price $r$.
Let $H$ be the set of all households, indexed by $i=1,\ldots,N$, and let $O$ be the set of owners.
Then for any given rental price $r$, the total rental demand is $\sum_{i\in H \setminus O}y_i^\star(r)$, and the total rental supply is $\sum_{i\in O}(\overline{y}_i-y_i^\star(r))$.
The market-clearing rental price $\tilde{r}$ is that which equates rental supply and demand.
The market-clearing rental volume $v$ is equal to the total rental supply (and the total rental demand) at rental price $\tilde{r}$.

Let $w_i$ be the surplus a household enjoys from participating in the rental market.
If household $i$ owns the asset, $w_i=f_i(y_i^\star(\tilde{r}))+\tilde{r}(\overline{y}_i-y_i^\star(\tilde{r}))-f_i(\overline{y}_i)$.
If household $i$ doesn't own the asset, $w_i=f_i(y_i^\star(\tilde{r}))-\tilde{r}y_i^\star(\tilde{r})$.
The total owner surplus is the sum of $w_i$ across all owner households, and the total renter surplus is the sum across all non-owner households.
The total surplus across all households is $\sum_{i\in H}f_i(y_i^\star)-\sum_{i\in O}f_i(\overline{y}_i)$, with the rent payments canceling out.

\subsection{Adoption level and pattern}

The rental market equilibrium depends on the adoption level and pattern.
Let $t$ be the percentage of households that adopt and thus own the DER asset.
For a given level of adoption $t$, define the adopters $O(t)$ as the set of households in the top $t$ percent when ordered by $\frac{f_i(\overline{y}_i)}{\overline{y}_i}$.
This adoption pattern corresponds to the notion that individual households will decide whether to own the DER asset based on how much they stand to save if they use it entirely for themselves, without considering possible revenue from the peer-to-peer rental market.
\footnote{This is an \emph{a-priori} adoption decision as formulated in \cite{horton2016}.}
They normalize their potential bill savings $f(\overline{y}_i)$ by their system size $\overline{y}_i$ to come up with a per-unit savings that they can compare directly to a per-unit purchase price.
Define the rate demand function $D_R(p)$ as the rate of adoption that corresponds to a per-unit asset purchase price of $p$; that is,
\begin{equation}
D_R(p)=t: \forall i\in O(t),\, \frac{f_i(\overline{y}_i)}{\overline{y}_i}\geq p.
\end{equation}

Define the quantity demand function $D(p)$ as the total quantity adopted at level $t$: 
\begin{equation}
D(p)=\sum_{i\in O(D_R(p))}\overline{y}_i.
\end{equation}

We will write $\tilde{r}(t)$ to refer to the market-clearing per-unit rental price when the adopters are $O(t)$, and $v(t)$ to refer to the corresponding market-clearing rental volume.

\subsection{Long run equilibrium}

A long run equilibrium of the peer-to-peer market is a point at which households are indifferent between renting and owning the asset.
In other words, it is a point at which the per-unit purchase price of the DER asset is the same as the per-unit rental market price. 
Suppose the per-unit purchase price $p$ is fixed.
Then the long run equilibrium adoption rate $\overline{t}$ satisfies $\tilde{r}(\overline{t})=p$.
Define the long run rate demand function $\tilde{D}_R(p)$ as:
\begin{equation}
\tilde{D}_R(p) = \overline{t}:\tilde{r}(O(\overline{t}))=p.
\end{equation}

Conceptually, for a fixed $p$, interpret $D_R(p)$ as the adoption level in the absence of the peer-to-peer market.
Once the peer-to-peer market comes into being, if the rental market price $\tilde{r}(D_R(p))$ is higher than $p$, there is profit to be made in purchasing the asset just to rent it out.
Once all of those opportunities are taken, the adoption level will be $\tilde{D}_R(p)$, under the assumption that the additional adoption comes from households who have the highest values of normalized bill savings.
Define the long run quantity demand function $\tilde{D}(p)$ as the total quantity adopted in the long run in the presence of the peer-to-peer market:
\begin{equation}
\tilde{D}(p)=\sum_{i\in O(\tilde{D}_R(p))}\overline{y}_i.
\end{equation}

\subsection{Geographical localness of market}
\label{dispersion}

The peer-to-peer rental market model assumes that the DER asset capacity can be freely transferred between households, without considering the requirements of transmission and distribution.
The system operator or the utility may be concerned with what these transfers look like --- whether they take place relatively locally or span long distances.
We develop two metrics to assess to what extent a given rental market is localized.
Let the geographic area containing all the households be partitioned into regions.
The set of all regions is $Z=\lbrace z_1,\ldots,z_{|Z|}\rbrace$.
Each household belongs to one and only one region.
Let $m_k$ be the set of all households that belong to region $z_k$.

For a given adoption level $t$, the set of adopters is $O(t)$, the peer-to-peer market rental price is $\tilde{r}(t)$, and the quantity rented is $v(t)$.
Define the excess rental supply in region $z_k$ as:
\begin{equation*}
s_k=\bigg(\sum_{i \in m_k \cap O(t)}\overline{y}_i-y_i^\star\big(\tilde{r}(t)\big)\bigg)-\sum_{i \in m_k \setminus O(t)}y_i^\star\big(\tilde{r}(t)\big).
\end{equation*}
If $s_k<0$ then region $z_k$ has excess rental demand.
The distance between regions $z_k$ and $z_l$ is $d(z_k,z_l)$.
Let $W$ be a $|Z|\times|Z|$ matrix with nonnegative entries, where $W_{kl}$ is the flow of rental supply from region $z_k$ to region $z_l$.
Define the minimum cost flow problem for rental market clearing as follows:

\begin{subequations}
\label{eqn:MincostFlow}
\begin{align}
\underset{W\geq 0}{\text{minimize}} \; & \sum_{k=1}^{|Z|} \sum_{l=1}^{|Z|} W_{kl}d(z_k,z_l)^2\\
\mbox{subject to} \; & \sum_{l=1}^{|Z|}W_{kl}=\textrm{max}(0,s_k),\; \forall k \in \lbrace1,\ldots,|Z|\rbrace\\
& \sum_{l=1}^{|Z|}W_{lk}=\textrm{max}(0,-s_k),\; \forall k \in \lbrace1,\ldots,|Z|\rbrace.
\end{align}
\end{subequations}

Constraints \ref{eqn:MincostFlow}(b) and (c) ensure that the inter-region flows match excess rental supply with excess rental demand and that no region acts as a simple passthrough circumventing the squared distance cost.
The minimum cost flow objective is a measure of how hard it is to match excesses with shortages taking into account the distance between regions.
Another useful measure is the fraction of total rental market quantity that can be matched, or cleared, within regions: $1-\frac{1}{2v(t)}\sum_{k=1}^{|Z|}|s_k|$.
Both of these measures are associated with the outcome of a peer-to-peer market in which all households can trade with all other households regardless of what region they're in.

In general, all regions will have some nonzero excess rental supply, so the minimum cost flow objective will be positive, and the fraction of market quantity cleared locally will be less than 1.
If by chance the rental quantity supplied exactly matches the rental quantity demanded within each region, then the excess supply in all regions is zero, so the minimum cost flow objective is zero, and the fraction of rental market quantity that can be cleared within regions is 1.

\subsection{Opposition between utility and equipment vendor}

The retail utility and DER equipment vendor have vested interests in whether or not the peer-to-peer market exists.
The retail utility may oppose the peer-to-peer market because it would lead to lower revenue as customers' electric bills decrease.
Furthermore, the utility may also oppose the market because it believes the market will increase the cost and complexity of operating the distribution grid.
By contrast, equipment vendors have a direct interest in the peer-to-peer market because it could lead to more purchases of the DER asset.
We model the interaction between the utility and the vendor as it relates to blocking or enabling the market.
We assume that the vendor is the agent seeking to open the peer-to-peer market, so we limit ourselves to cases where it stands to gain from doing so.

Suppose there is a single equipment vendor.
Let $D(p)$ be the quantity demand curve the vendor faces in the absence of the peer-to-peer market, and $\tilde{D}(p)$ the long run quantity demand curve in the presence of the market.
Suppose the vendor sells the DER asset at a per-unit price $p$ which does not change depending on the existence of the peer-to-peer market.
Without the peer-to-peer market, the vendor's revenue is $pD(p)$; with the market, it is $p\tilde{D}(p)$.
The only difference between the two situations is the quantity sold.
In total, the peer-to-peer market increases the vendor's revenue by $\Delta R_V(p)=p(\tilde{D}(p)-D(p))$.

Now we turn to the utility.
Suppose the utility earns a profit on the amount it bills for energy sold to a household, but does not earn profit or incur loss when it buys back electricity from households.
Thus the utility desires to sell as much energy as possible, and it doesn't care about how much it buys back.
Let $B_e(q)$ be total billed sales without the peer-to-peer market when the adopted quantity is $q$, and $\tilde{B}_e(q)$ be total billed sales with the peer-to-peer market.
If the adopted quantity is $q$ without the market and $\tilde{q}$ with the market, then the utility's loss of billed sales due to the peer-to-peer market is $\Delta R_U(q,\tilde{q})=B_e(q)-\tilde{B}_e(\tilde{q})$.
For a fixed purchase price $p$, $q=D(p)$, and $\tilde{q}=\tilde{D}(p)$, so $\Delta R_U(q,\tilde{q})=\Delta R_U(p)=B_e(D(p))-\tilde{B}_e(\tilde{D}(p))$.

Let the utility's rate of profit on its billed sales revenue be fixed at $\zeta_U$, and let the vendor's rate of profit on its revenue be fixed at $\zeta_V$.
More precisely, $\zeta_U$ and $\zeta_V$ are constant, gross rates of profit on the increase or decrease of revenue due to the existence of the peer-to-peer market.
The vendor's profit increase from the market is $\zeta_V\Delta R_V(p)$, and the utility's profit loss is $\zeta_U\Delta R_U(p)$.
These potential profit gains or losses represent resources that the vendor or utility may bring to bear to enable or block the emergence of the market.
We assume that the peer-to-peer market will come into existence if the vendor stands to gain more profit than the utility will lose, i.e. if:
\begin{equation}
\frac{\Delta R_V(p)}{\Delta R_U(p)}>\frac{\zeta_U}{\zeta_V}
\end{equation}
Define the utility's profit rate ratio as $A_U=\zeta_U/\zeta_V$.
For a given $p$, if $A_U$ is high enough, the utility will win out and block the emergence of the peer-to-peer market.

\subsection{Comparison with direct subsidies}

\begin{figure}
\centering
\includegraphics[width=.6\textwidth]{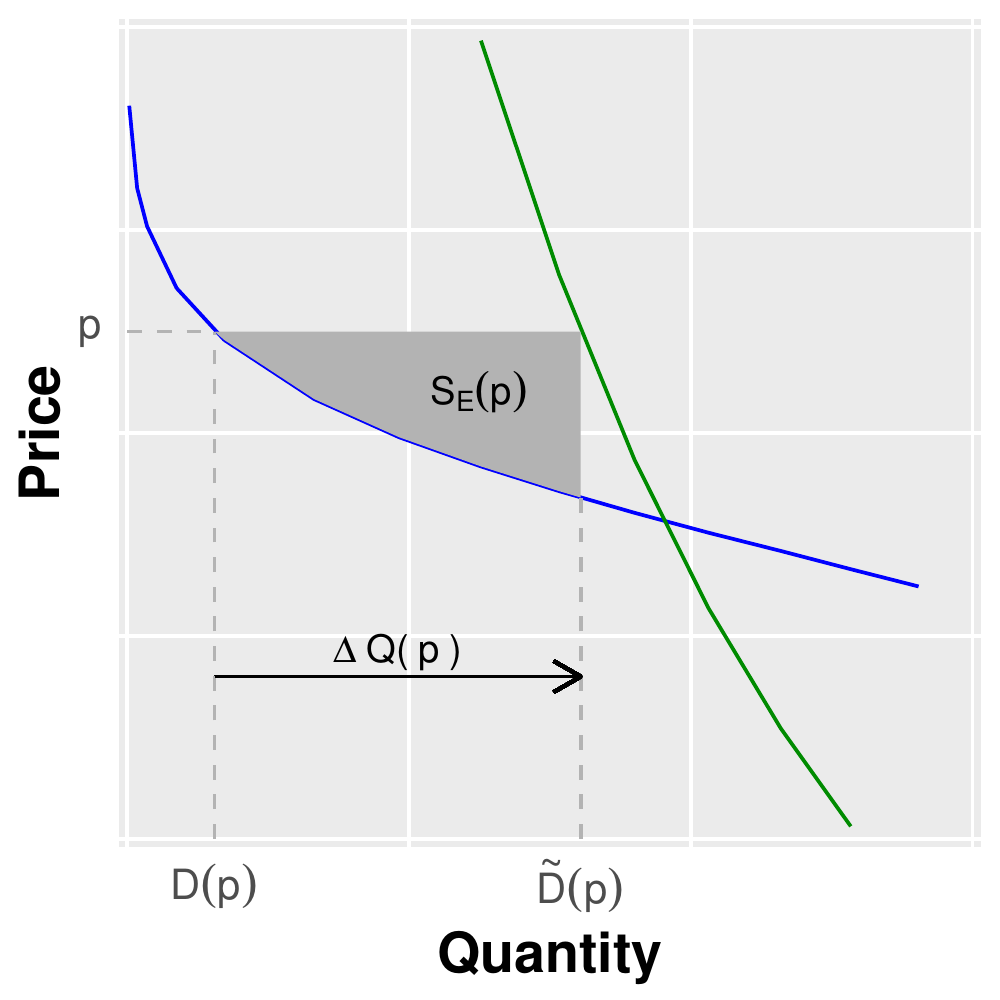}
\caption{This figure defines the equivalent direct subsidy of the peer-to-peer market.
The blue curve is the demand curve without the market, and the green curve is the long run demand curve in the presence of the market.
For a fixed asset purchase price $p$, adoption without the market is $D(p)$ and with the market is $\tilde{D}(p)$.
The difference between the two is $\Delta Q(p)$.
The direct subsidy that generates the same increase in adopted quantity is $S_E(p)$.}
\label{fig:subsidy_diagram}
\end{figure}

For a range of asset purchase prices, the emergence of the peer-to-peer market would increase total asset adoption because the long run rent-own equilibrium involves a higher level of adoption.
The increase in total adopted quantity due to the market is $\Delta Q(p)=\tilde{D}(p)-D(p)$.
Direct subsidies are an alternate way to increase adoption.
We define the equivalent direct subsidy $S_E(p)$ for a fixed asset purchase price $p$ as the amount of subsidy required to increase the adopted quantity by the same amount as the peer-to-peer market would.
We compute it as $S_E(p)=\int_{D(p)}^{\tilde{D}(p)}(p-D^{-1}(q))dq$.
Figure \ref{fig:subsidy_diagram} illustrates these quantities.

\section{Data}

The household inflexible loads come from the smart meter data for over 45,000 households in and around Fresno, California, for one year spanning from November 2011 to October 2012.
We do not include meters with very low consumption (less than $0.1$ kW mean) or many zero readings (more than half of all readings).
The zip code for each household is available.
The peer-to-peer market involves renting \emph{capacity}; therefore, the households included in the market should experience very similar solar irradiance so that the PV component of the DER asset is truly fungible.\footnote{The slope coefficient for an ordinary least squares regression between the solar irradiance for any two zip codes for this set of households ranges from 0.96 to 1.00. The maximum mean absolute difference between any two zip codes is 26 Wh/$\textrm{m}^2$.}

We follow \cite{patel2017} in obtaining solar irradiance data.
To remove any variation between households due to irradiance variations, we use a single solar irradiance for all households.
We choose the irradiance from the zip code that is closest to the weighted centroid of all zip codes for this set of households, where each zip code is weighted by the total annual electricity consumption of the households in it.

We use Policy 1 pricing from \cite{patel2017}: households buy electricity at the retail time of use rate and sell surplus electricity back to the grid at the day ahead location marginal price (LMP).
The LMP varies by zip code, which is an extraneous source of variation for this study.\footnote{The slope coefficient for an ordinary least squares regression between the LMPs for any two zip codes for this set of households around Fresno ranges from 0.77 to 1.15. The maximum mean absolute difference between any two zip codes is 0.2 \textcent/kWh.}
Therefore, we generate a single LMP for all households by taking the mean of the zip code LMPs weighted by total annual household electricity consumption in each zip code.

We follow \cite{patel2017} in defining and sizing the DER asset, which is a combination of a PV system and a storage device, and its interface to the household AC bus.
The PV system is sized to be net-zero using the irradiance described in the preceding paragraph.
The storage device capacity, charging rate, and discharging rate all scale to the PV system size, with $\alpha=1$ kWh/kW, and $\overline{u}=\underline{u}=\frac{5}{13.5}$ kWh/kW.

\subsection{Savings functions}

For each household, we solve Optimization (\ref{eqn:Optimization}) for each day of the year, obtaining yearly savings for 30 values of $y_i$ ranging from $0.01\overline{y}_i$ to $\overline{y}_i$.
We then fit a monotonically increasing, concave spline to this data to generate $f_i(y_i)$ \cite{cobsarticle,cobsmanual}.

\begin{figure}
\centering
\begin{subfigure}[b]{0.475\textwidth}
\centering
\includegraphics[width=\textwidth]{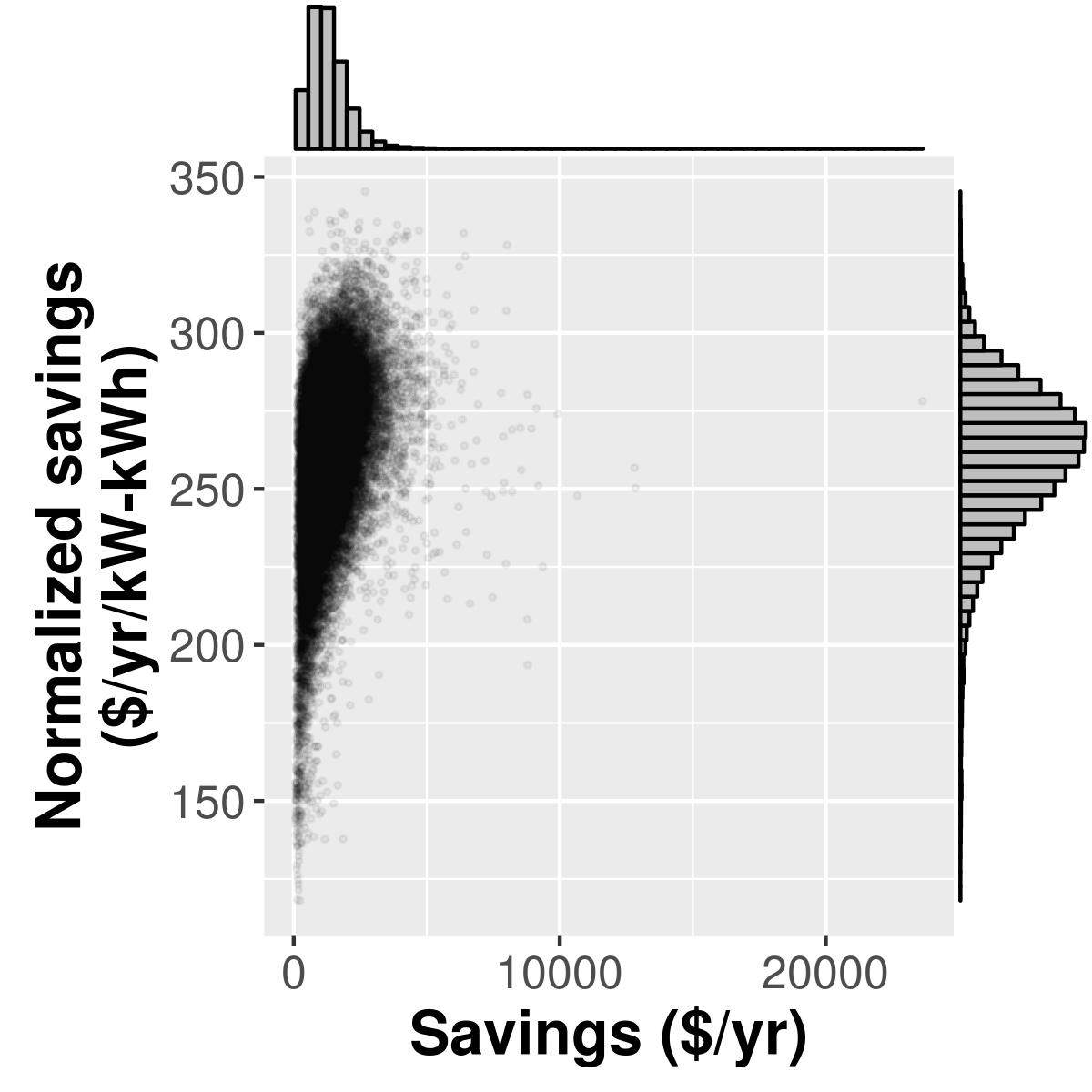}
\caption{}
\label{fig:f1_dist}
\end{subfigure}
\hfill
\begin{subfigure}[b]{0.475\textwidth}  
\centering 
 \includegraphics[width=\textwidth]{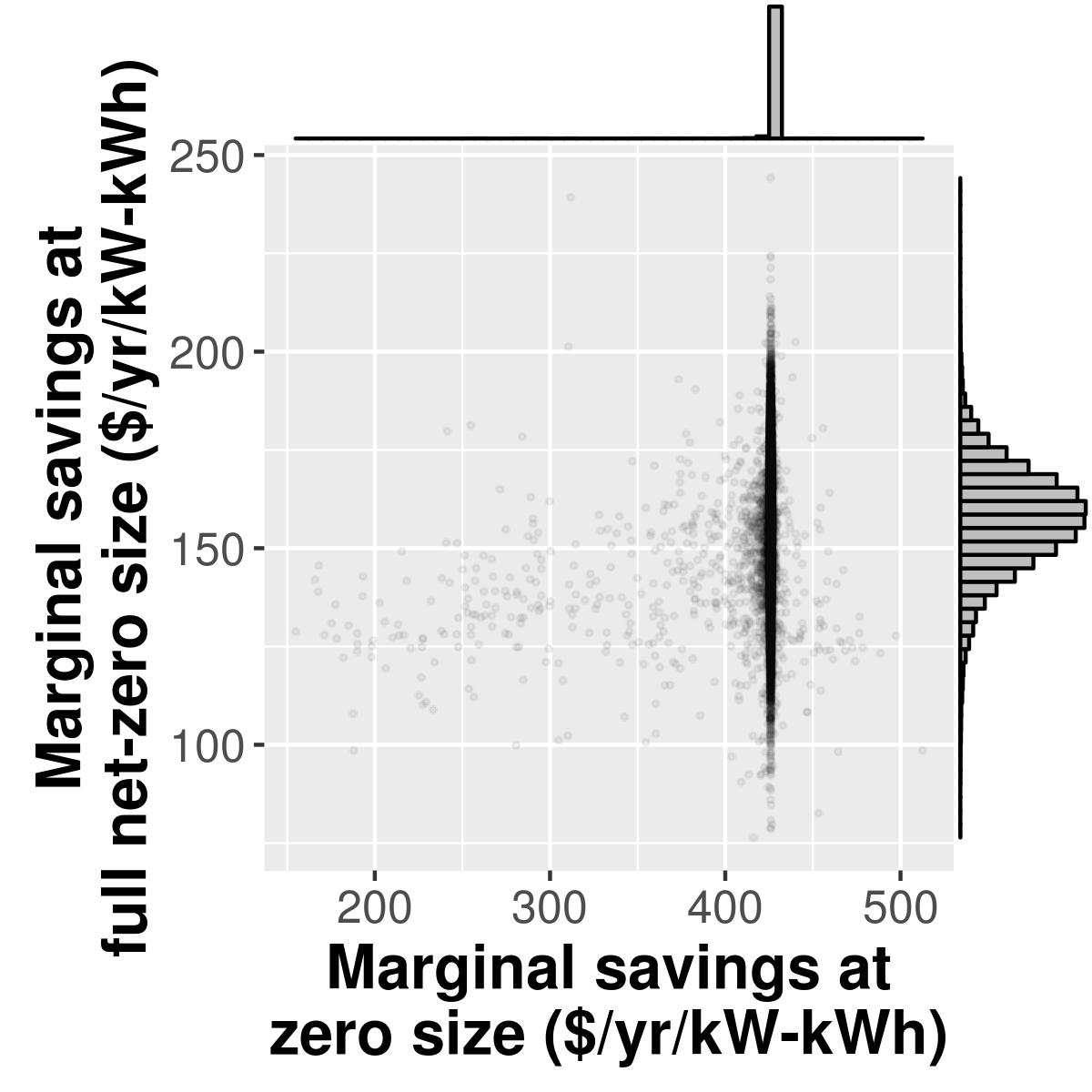}
\caption{}
\label{fig:fp1_v_fp0}
\end{subfigure}
\vskip\baselineskip
\begin{subfigure}[b]{0.475\textwidth}   
\centering 
 \includegraphics[width=\textwidth]{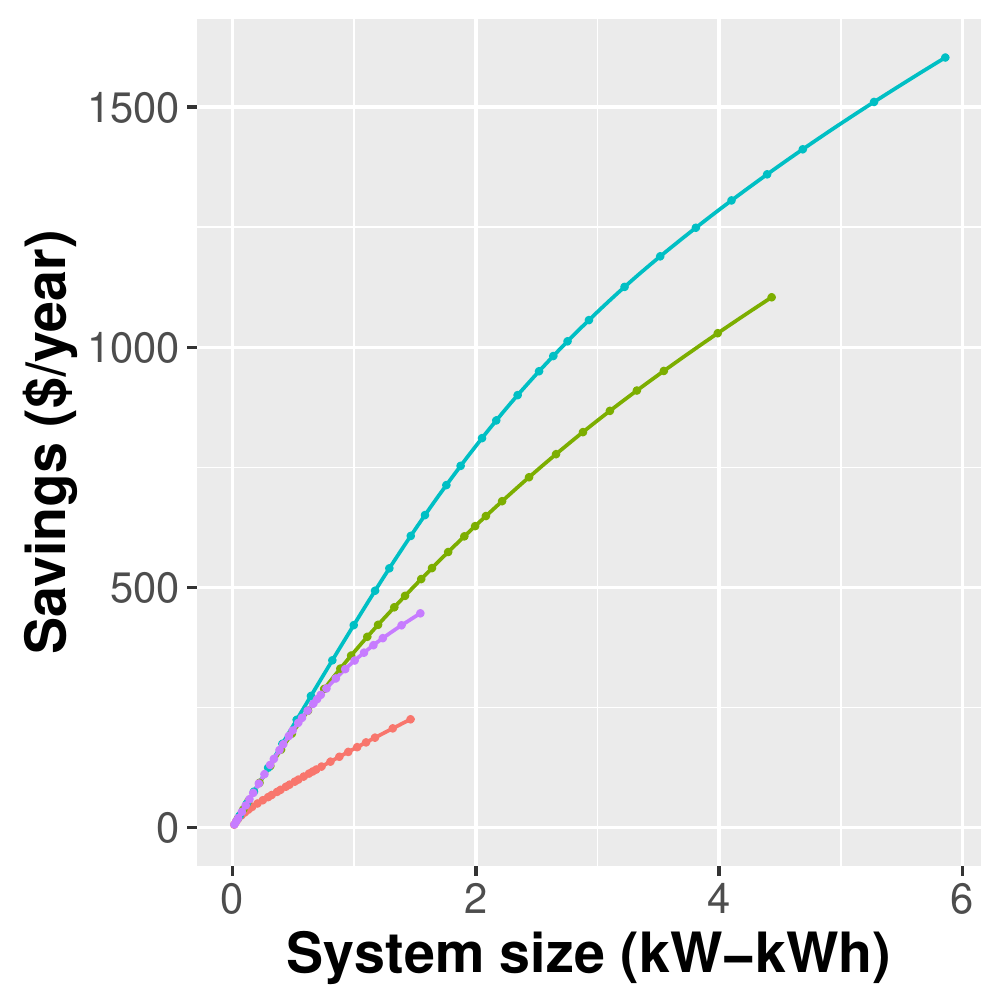}
\caption{}  
\label{fig:f_plots}
\end{subfigure}
\quad
\begin{subfigure}[b]{0.475\textwidth}   
\centering             \includegraphics[width=\textwidth]{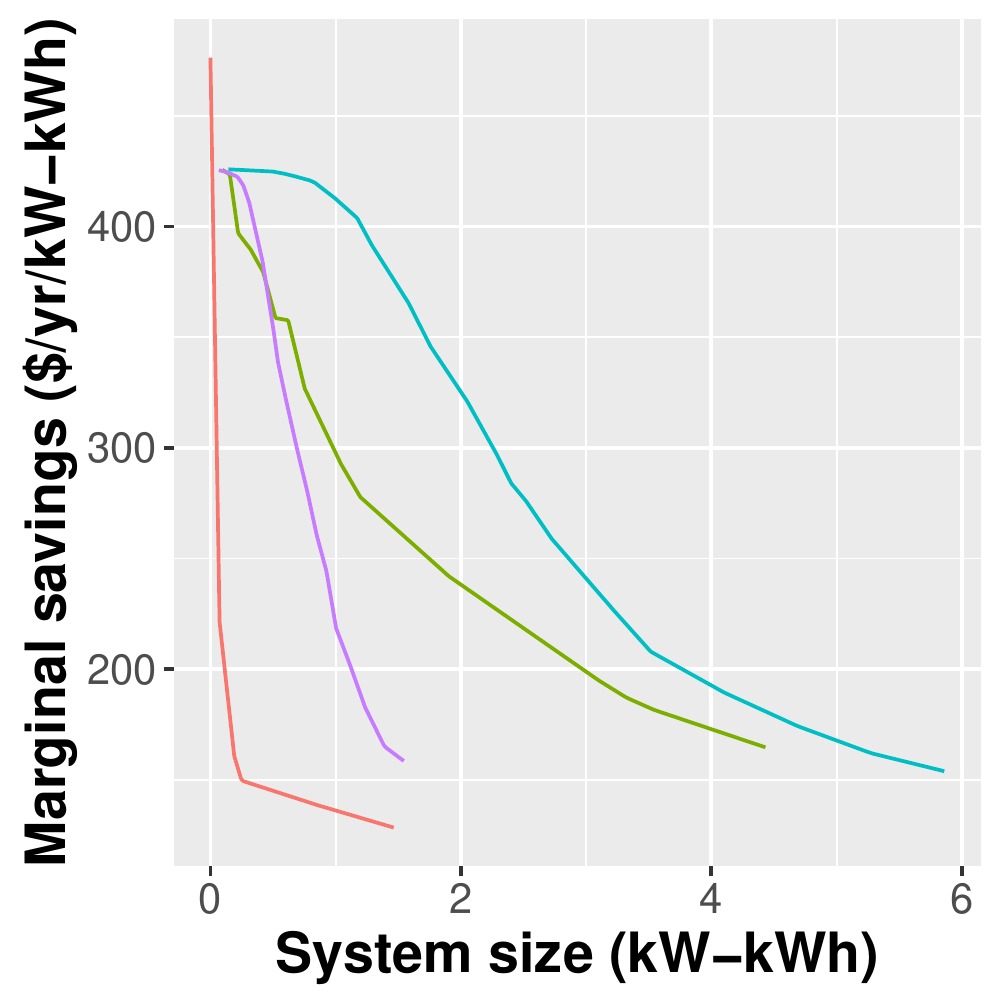}
\caption{}   
\label{fig:fp_plots}
\end{subfigure}
\caption{The savings function for a household $f_i$ is a spline fitted to 30 points computed by varying $y_i$ in Optimization (\ref{eqn:Optimization}). (a) If a household owns a net-zero sized system and uses it for itself, it achieves absolute annual savings $f_i(\overline{y}_i)$ and normalized savings $f_i(\overline{y}_i)/\overline{y}_i$. The two values are plotted against each other with marginal histograms. The absolute savings and normalized savings are moderately correlated. (b) The initial marginal savings ($f_i'(0)$) and the final marginal savings at full net-zero system size ($f_i'(\overline{y}_i)$) are plotted against each other with marginal histograms.
(c) The savings functions for four example households are plotted, with the dots showing the values computed from Optimization (\ref{eqn:Optimization}).
The splines fit the values very well, without noticeable over-fitting.
The red curve is for the household with the worst fit; its $\textrm{R}^2$ exceeds $1-10^{-5}$. (d) The derivative of the savings function is plotted for the same four example households.} 
\label{fig:Savings_function_stats}
\end{figure}

Figure \ref{fig:Savings_function_stats} illustrates the heterogeneity in key parameters of the savings functions across the households.
The normalized savings that a household can obtain by owning the asset and using it entirely itself ranges from about \$125-325/yr/kW-kWh.\footnote{Normalized savings are computed as $f(\overline{y}_i)/\overline{y}_i$, but because the asset includes a storage device whose capacity is 1 kWh for every 1 kW of PV capacity, we report the normalized savings as \$/yr/kW-kWh.}
As seen in the marginal histograms in Figure \ref{fig:fp1_v_fp0}, the initial slope of the savings function is almost the same for all households, while the terminal slope has some variation.
The reason for this is as follows.
All households face the same TOU rate and experience the same irradiance.
Furthermore, most households have non-zero consumption at all times.
Thus, with the first increment of DER capacity, most households apply the same small amount of PV energy generated to offset their own consumption.
That consumption is priced the same for all households, so they all save about the same amount for very small values of $y_i$.
As the DER asset capacity grows, households who don't have as much peak period consumption will then apply the capacity to offset off-peak consumption, or sell back surplus generation to the grid, both of which yield smaller benefits.
Therefore, the variation in consumption patterns across households will lead to variation in savings as $y_i$ increases.
Figures \ref{fig:f_plots} and \ref{fig:fp_plots} give a few examples of the savings data and the fitted functions, demonstrating that the spline fits are excellent.

\section{Results}

\subsection{Rental market characteristics}

For adoption rates ranging from $t=0.1\%$ to $99.9\%$, the peer-to-peer market rental price $\tilde{r}$ and quantity $v$ are computed.
A participant is a household that either rents or rents out a positive fraction of the DER asset on the market at equilibrium.
Formally, at a given equilibrium, household $i$ is a participant in the market if it is a non-owner and $y_i^\star>0$ or it is an owner and $y_i^\star<\overline{y}_i$.

\begin{figure}
\centering
\begin{subfigure}[b]{0.475\textwidth}
\centering
\includegraphics[width=\textwidth]{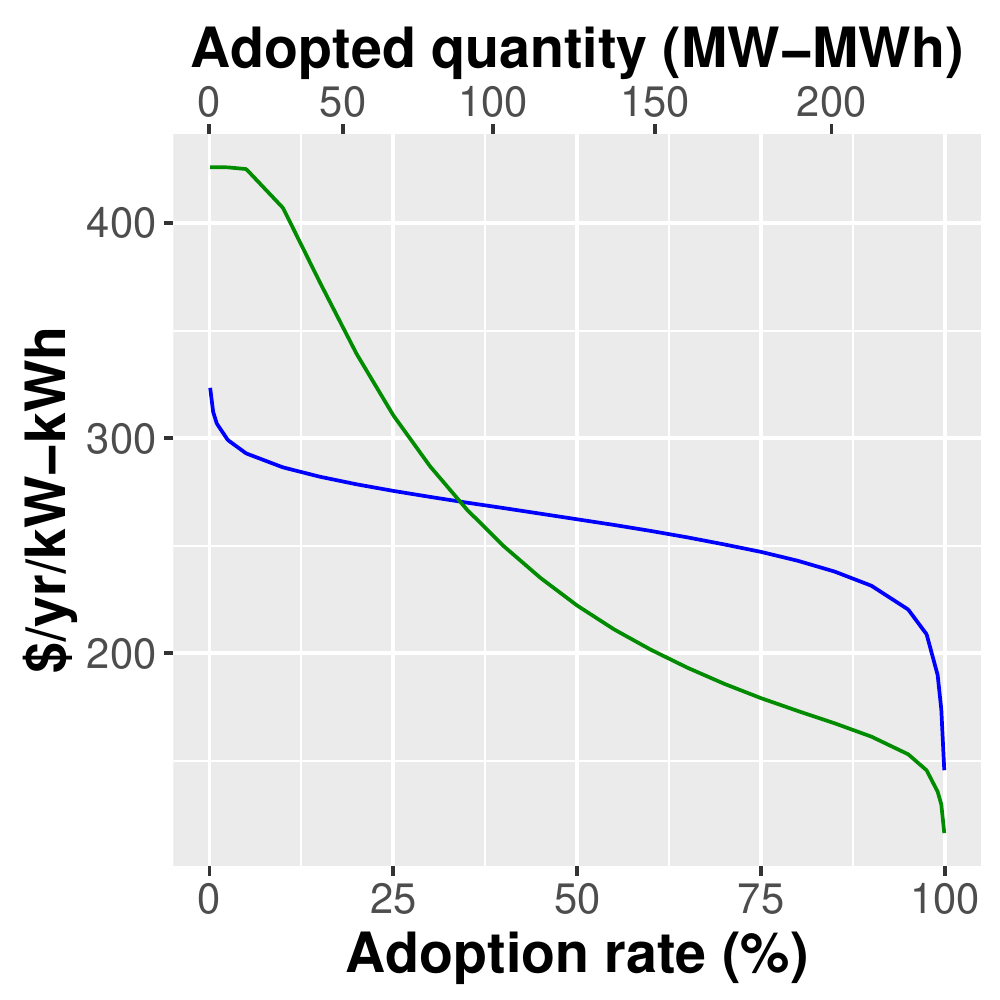}
\caption{}
\label{fig:long_run_eq}
\end{subfigure}
\hfill
\begin{subfigure}[b]{0.475\textwidth}  
\centering 
 \includegraphics[width=\textwidth]{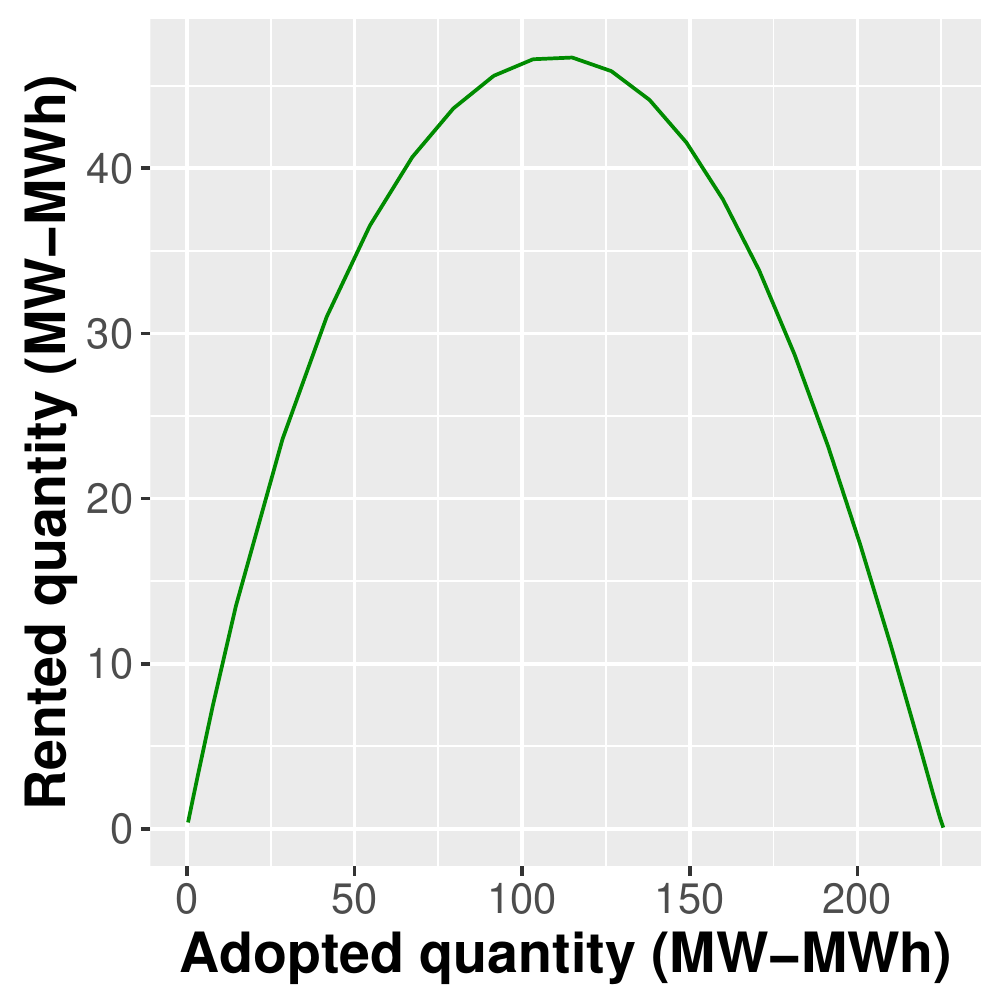}
\caption{}
\label{fig:q_star}
\end{subfigure}
\vskip\baselineskip
\begin{subfigure}[b]{0.475\textwidth}   
\centering 
 \includegraphics[width=\textwidth]{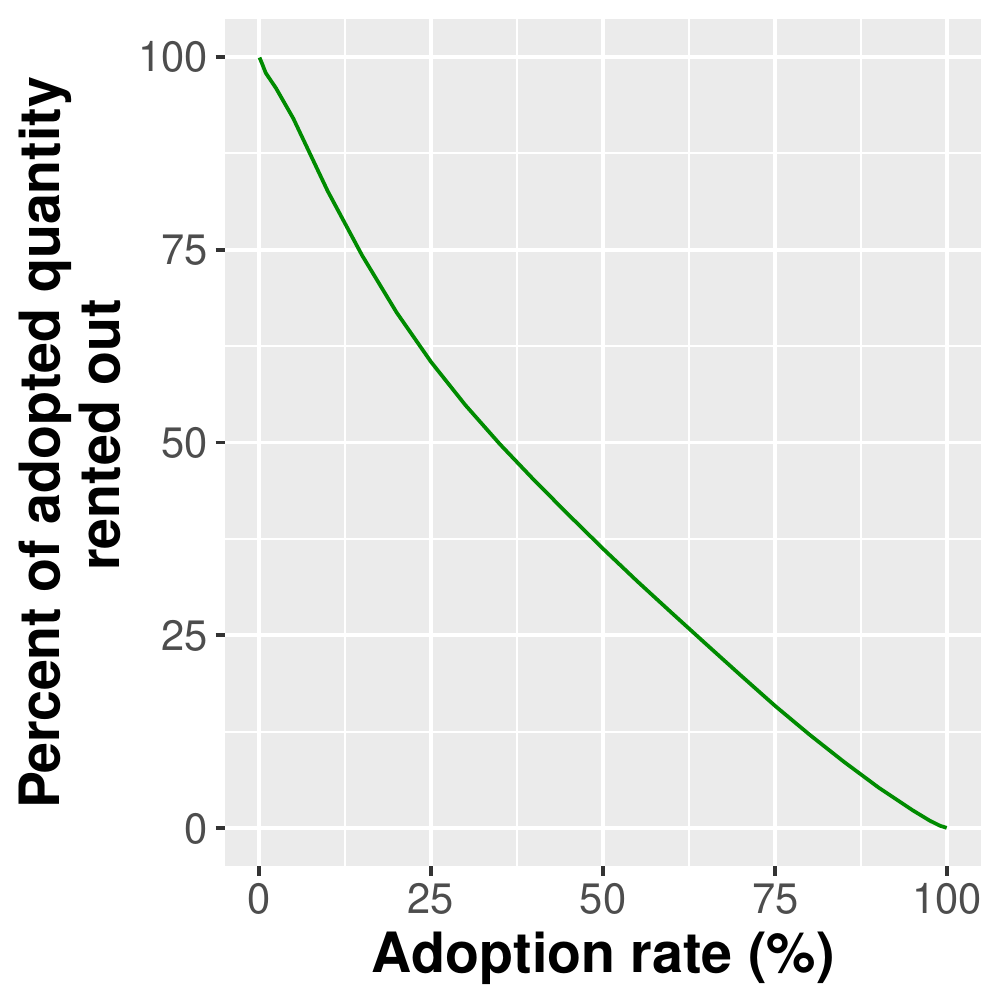}
\caption{}  
\label{fig:fraction_rented_out}
\end{subfigure}
\quad
\begin{subfigure}[b]{0.475\textwidth}   
\centering             \includegraphics[width=\textwidth]{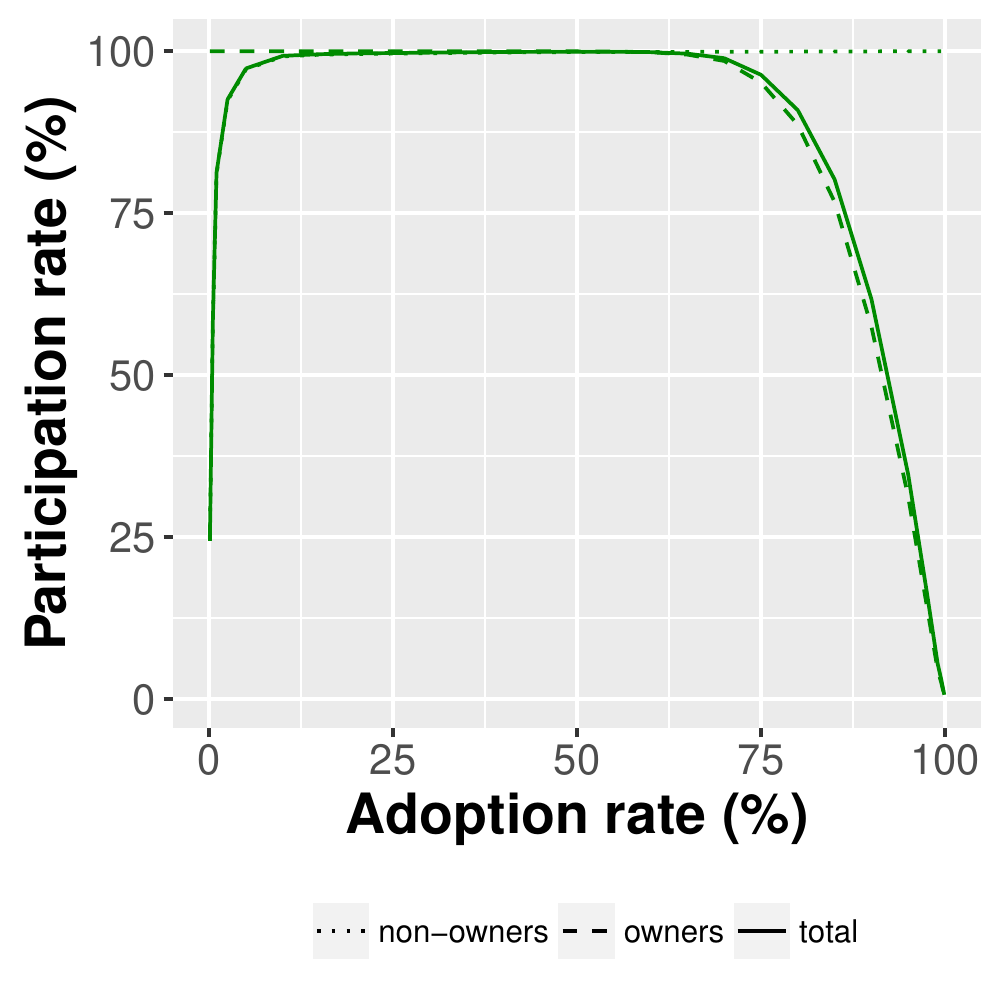}
\caption{}   
\label{fig:participation}
\end{subfigure}
\caption{(a) The blue curve is $D(p)$, the demand in the absence of the peer-to-peer market. The green curve gives the market rental rate for a given level of adoption.
The latter can also be interpreted as $\tilde{D}(p)$, the long run demand in the presence of the peer-to-peer market given an asset purchase price $p$. The x-axes give either the rate of adoption or the quantity adopted, allowing a comparison of the two.
(b) The quantity rented on the peer-to-peer market is plotted as a function of the total adopted quantity.
The maximum rental quantity occurs when total adoption is about 115 MW-MWh, which corresponds to around 45\% of households adopting.
(c) The percent of the adopted quantity rented out is plotted as a function of the adoption rate.
This percent decreases with increasing adoption. (d) Market participation rates are plotted as a function of adoption rate.
Between 10\% and 60\% adoption, almost all households participate in the peer-to-peer market.} 
\label{fig:p_q_fraction_participation}
\end{figure}

The rental price decreases at a fairly steady rate with increasing adoption, as shown in Figure \ref{fig:long_run_eq}.
At low levels of adoption, the market rental price is about \$420/yr/kW-kWh, which is well above the asset purchase price that produces the same level of adoption in the absence of the peer-to-peer market.
Households with larger net-zero systems tend to have higher normalized savings, which manifests in the moderate positive correlation between absolute savings and normalized savings seen in Figure \ref{fig:f1_dist}.
Thus, the first adopters tend to have larger net-zero sizes, which can be seen by comparing the adopted quantity axis with the adoption rate axis in Figure \ref{fig:long_run_eq}.

The market quantity hits a maximum point, as shown in Figure \ref{fig:q_star}.
At low adoption levels, the volume transacted is limited by the rate of adoption; at high adoption levels, almost everyone has the asset so there aren't that many renters available.
Figure \ref{fig:fraction_rented_out} illustrates that the fraction of all assets rented out steadily decreases with increasing adoption.
The maximum rented quantity occurs at about 45\% adoption, at which about 40\% of the available quantity of DER assets is rented out.

Figure \ref{fig:participation} shows how owner, non-owner, and total participation rates vary with adoption rate.
The participation rate is defined as the fraction of a given category that are participants in the market at equilibrium.
Participation across both categories is very high over a wide range of adoption rates, from 5\% to 75\%.

\subsection{Long run equilibrium}

Suppose the annualized purchase price of the DER asset is \$290/yr/kW-kWh.
The $D(p)$ curve in Figure \ref{fig:long_run_eq} indicates that about 5\% of households would choose to own the asset prior to the existence of the peer-to-peer market.
Now suppose the peer-to-peer market emerges.
The rental price on the market at that level of adoption is about \$425/yr/kW-kWh.
Thus, there is a profit opportunity in buying the asset just to rent it out.
In fact, the rental price is above \$290/yr/kW-kWh up until about 30\% adoption --- so another 25\% of households in addition to the original 5\% of adopters could buy the asset and make money on the rental market.
Once the total ownership rate hits 30\%, there is no longer any gain in owning to rent.
This is a long run rent-own equilibrium point.
In this example, the existence of the peer-to-peer market increases the overall ownership rate.

Consider a different scenario, with the annualized purchase price at \$250/yr/kW-kWh.
The adoption rate, prior to the existence of the peer-to-peer market, would be about 70\%.
When the peer-to-peer market comes into being, the rental rate would be about \$185/yr/kW-kWh.
The purchase price is greater than the rental rate, so adoption would not increase.
In fact, if possible, households who own the asset would choose to no longer be owners and instead to rent capacity on the peer-to-peer market, leading to a decrease in adoption.

\subsection{Comparison with coordination}

We compare the surplus from the peer-to-peer market with the value of coordinated action (VCA) as defined in \cite{patel2017}.
The VCA is the extra savings, in excess of what the owners of the asset save on their own, that is gained if all the households are aggregated into one household with one large load and one large DER asset controlled by a coordinator.
This type of coordination has a couple of advantages over the peer-to-peer market model.
For one, the coordinator can reallocate asset capacity on an hour by hour basis, whereas the peer-to-peer market reallocates capacity just once for the whole year in our study.
In addition, the peer-to-peer market can only shift one unit of DER asset capacity to or from a household (i.e., $y_i\in[0,\overline{y}_i]$), while the coordinator is not limited by that restriction.

\begin{figure}
\centering
\begin{subfigure}[b]{0.475\textwidth}
\centering
\includegraphics[width=\textwidth]{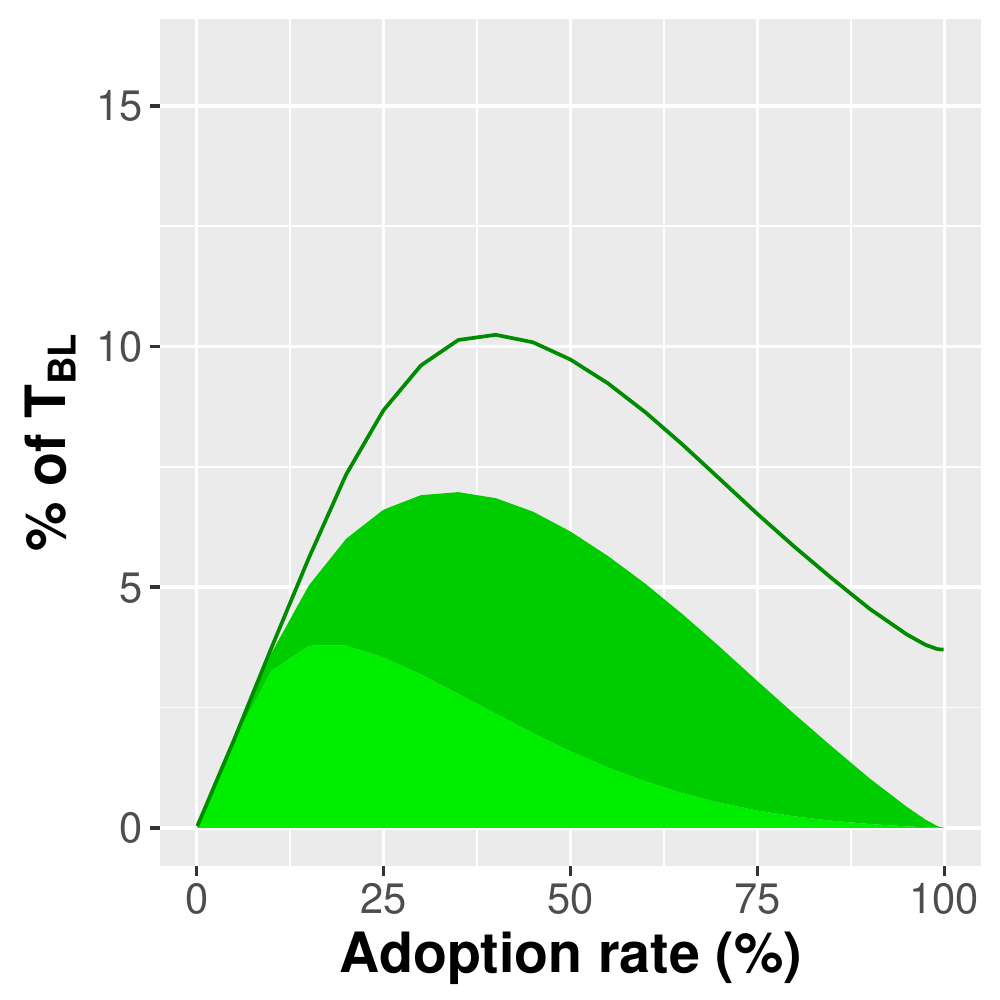}
\caption{}           \label{fig:VCA_surplus_fraction_baseline_fwd}
\end{subfigure}
\hfill
\begin{subfigure}[b]{0.475\textwidth} 
\centering 
 \includegraphics[width=\textwidth]{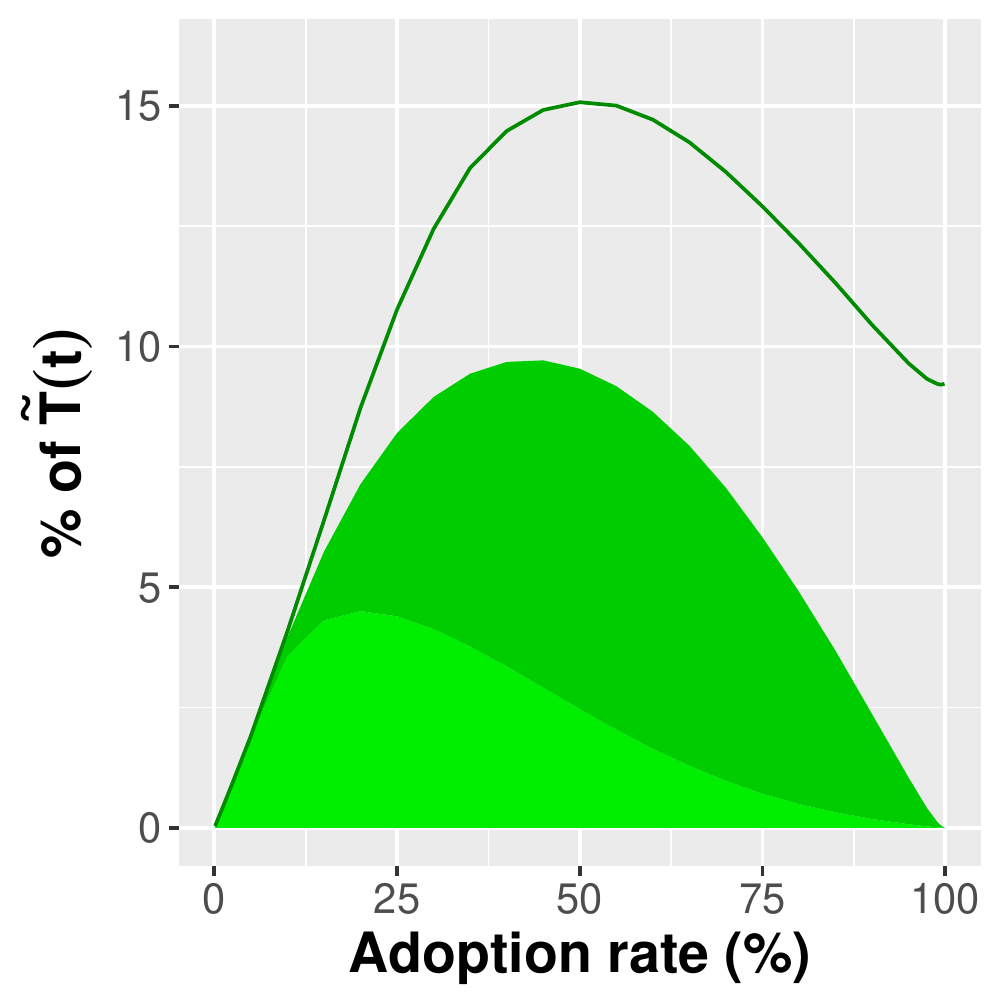}
\caption{}              \label{fig:VCA_surplus_fraction_preP2P_fwd}
\end{subfigure}
\caption{The solid line is the VCA. The shaded region is the total peer-to-peer market surplus, with the lighter region being the owner surplus and the darker region the renter surplus.
Fig. (a) uses $T_{BL}$, the original total cost of electricity for the households, as the normalizing quantity.
Fig. (b) uses $\tilde{T}(t)$, the total cost of electricity in the absence of the peer-to-peer market, as the normalizing quantity.} 
\label{fig:VCA_P2P_surplus}
\end{figure}

These two major advantages ensure that the VCA will exceed the total peer-to-peer market surplus, as it indeed does in Figure \ref{fig:VCA_P2P_surplus}.
Let $T_{BL}=\sum_{i\in H}b_{BL,i}$, the original total electricity bill for all households prior to any DER adoption.
The maximum VCA is about 10\% of $T_{BL}$, and it occurs at about 40\% adoption; the maximum total market surplus is about 7\% of $T_{BL}$ at 35\% adoption.
Interestingly, however, at lower levels of adoption, the peer-to-peer market generates almost as much surplus as the coordinator.
The gap between the two grows with increasing adoption up to a point.

Another insight is obtained by dividing the VCA and the surplus by $\tilde{T}(t)=\sum_{i\in O(t)}b_i(\overline{y}_i) + \sum_{i\in H\setminus O(t)}b_i(0) = T_{BL}-\sum_{i\in O(t)}f_i(\overline{y}_i)$.
$\tilde{T}(t)$ is the total amount paid to the utility by the households at adoption level $t$ when they cannot interact with each other through the market or through coordination.
This quantity changes with the adoption rate, in contrast to the constant $T_{BL}$.
As a percentage of $\tilde{T}(t)$, the VCA increases up to 15\% at about 50\% adoption, and the total market surplus peaks at 10\% at 40\% adoption.

\subsection{Localness of market}

In order to compute the measures of localness of the peer-to-peer market developed in Section \ref{dispersion}, we use zip codes as the regional partition.
There are 29 zip codes for the households in the Fresno area.
A representative latitude and longitude for each zip code serve as coordinates for defining inter-region distances $d(z_k,z_l)$.

\begin{figure}
\centering
\begin{subfigure}{.475\textwidth}
\centering
\includegraphics[width=\linewidth]{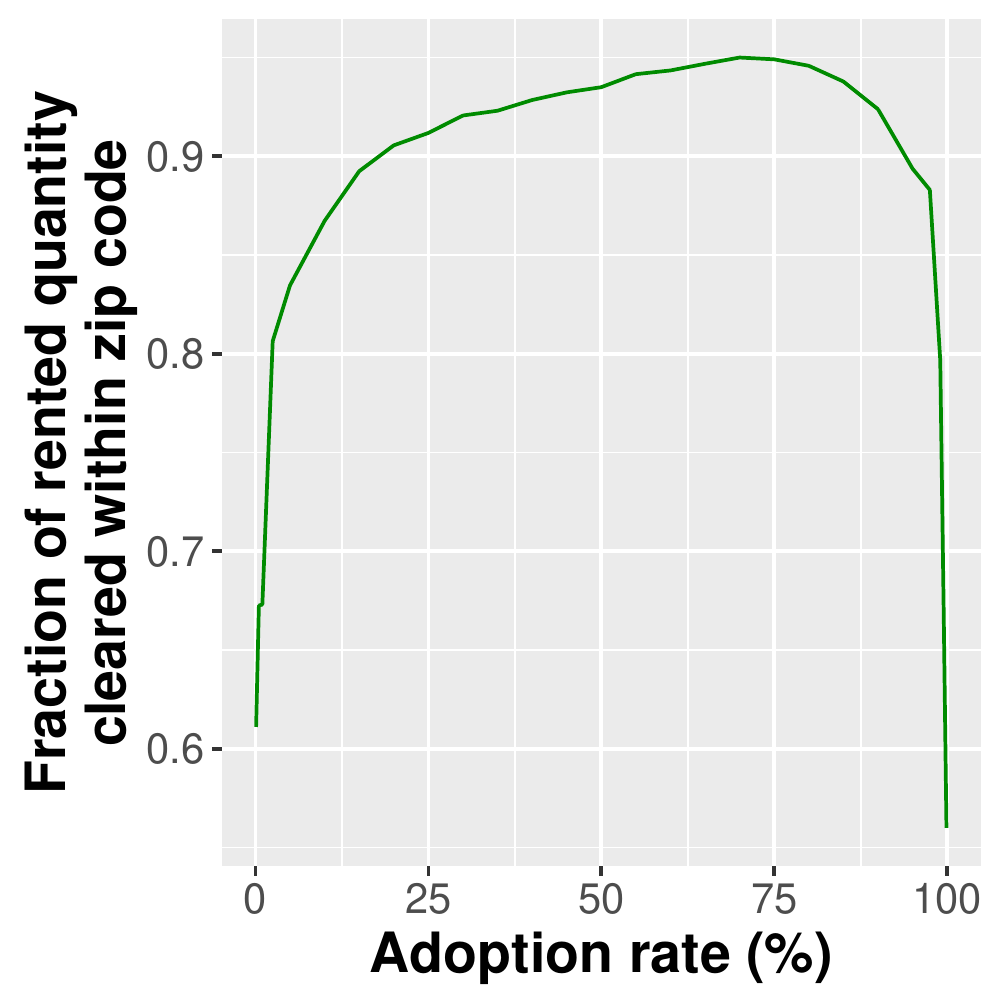}
\caption{}
\label{fig:fraction_q_star_cleared_in_zip}
\end{subfigure}%
\begin{subfigure}{.475\textwidth}
\centering
\includegraphics[width=\linewidth]{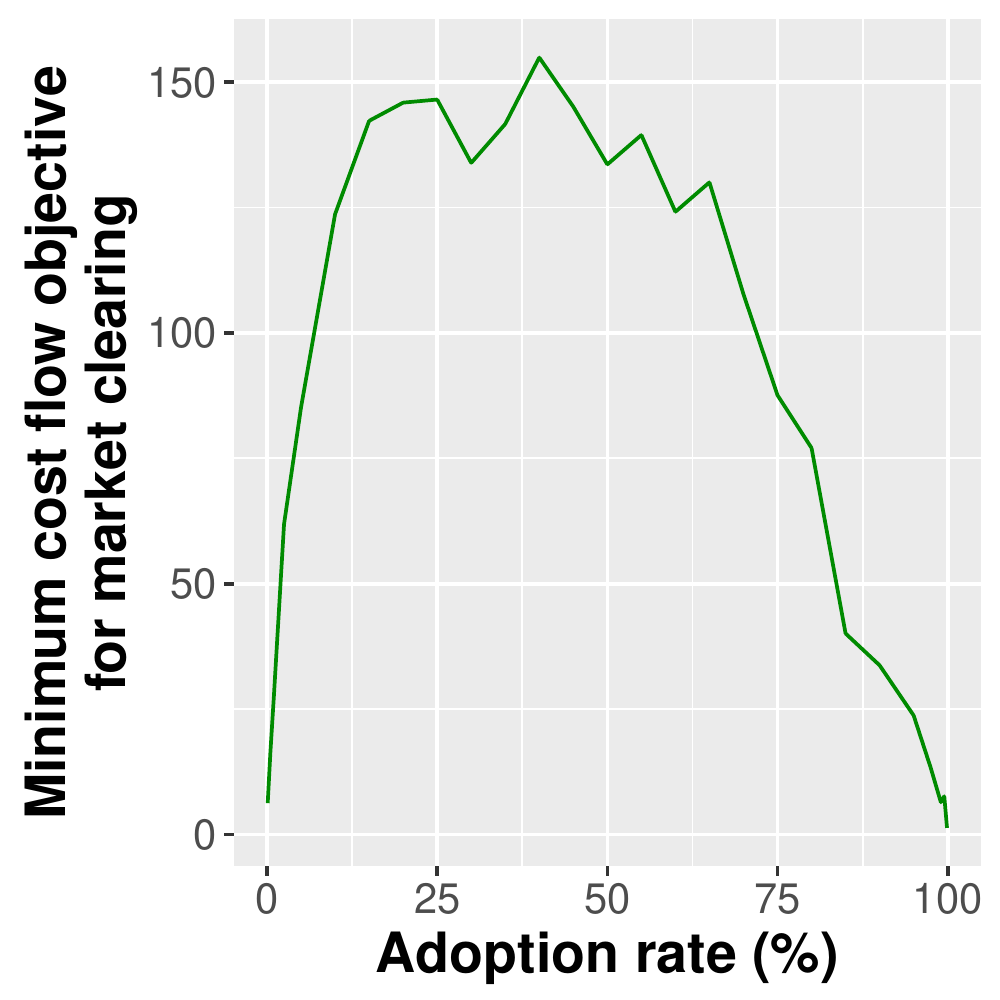}
\caption{}
\label{fig:min_cost_flow}
\end{subfigure}
\caption{(a) The curve gives the fraction of the rental quantity demand that can be satisfied by rental quantity supply from within the same zip code. For a wide range of adoption rates, over 90\% of rental market quantity can be cleared locally. (b) The objective for the minimum cost flow for market clearing is plotted as a function of DER adoption rate.}
\label{fig:q_star_frxn_min_cost_flow}
\end{figure}

As can be seen in Figure \ref{fig:fraction_q_star_cleared_in_zip}, for adoption rates between 15\% and 90\%, over 90\% of the rental market demand can be served by supply coming from within the same zip code.
This indicates a highly localized market for a broad range of adoption.
Even at very low and very high adoption rates, the majority of rental demand can be matched with supply within the same zip code. 
Thus, the peer-to-peer rental market is not dominated by large transfers of capacity between zip codes --- much of the market clears locally.

The minimum cost flow objective is plotted in Figure \ref{fig:min_cost_flow}.
When comparing this objective to the rented quantity as seen in Figure \ref{fig:q_star}, the major difference is that the objective is relatively high compared to the rented quantity in the range of 5-20\% adoption.
This is also the range over which the fraction of demand that can be matched within its own zip code is relatively low.
That mismatch between local rental supply and demand drives up the minimum cost flow objective for market clearing.
Interestingly, the objective is roughly the same from 20-60\% adoption, even though the rented quantity changes quite a bit over this range.
This confirms the finding that the market is well localized for a large range of adoption rates.

\subsection{Strategic interaction between utility and equipment vendor}

\begin{figure}
\centering
\begin{subfigure}{.475\textwidth}
\centering
\includegraphics[width=\linewidth]{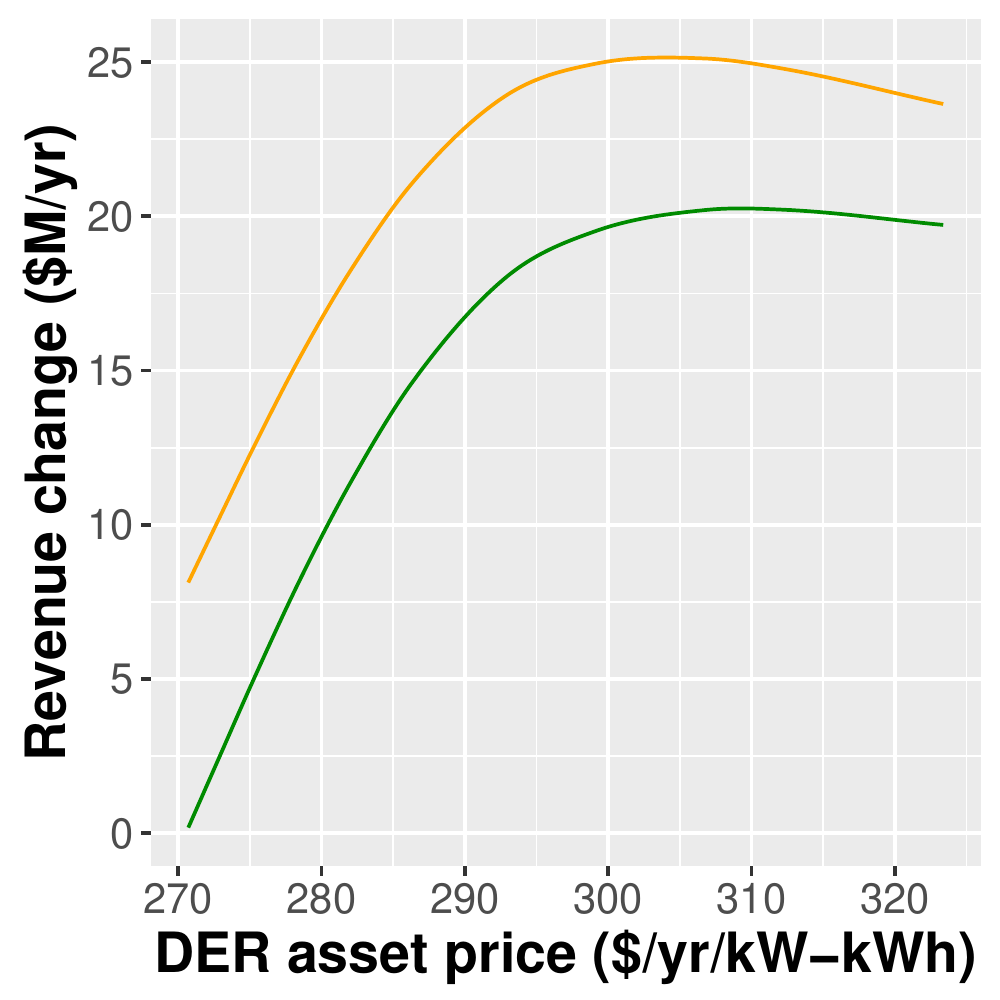}
\caption{}
\label{fig:fixed_price_DR}
\end{subfigure}%
\begin{subfigure}{.475\textwidth}
\centering
\includegraphics[width=\linewidth]{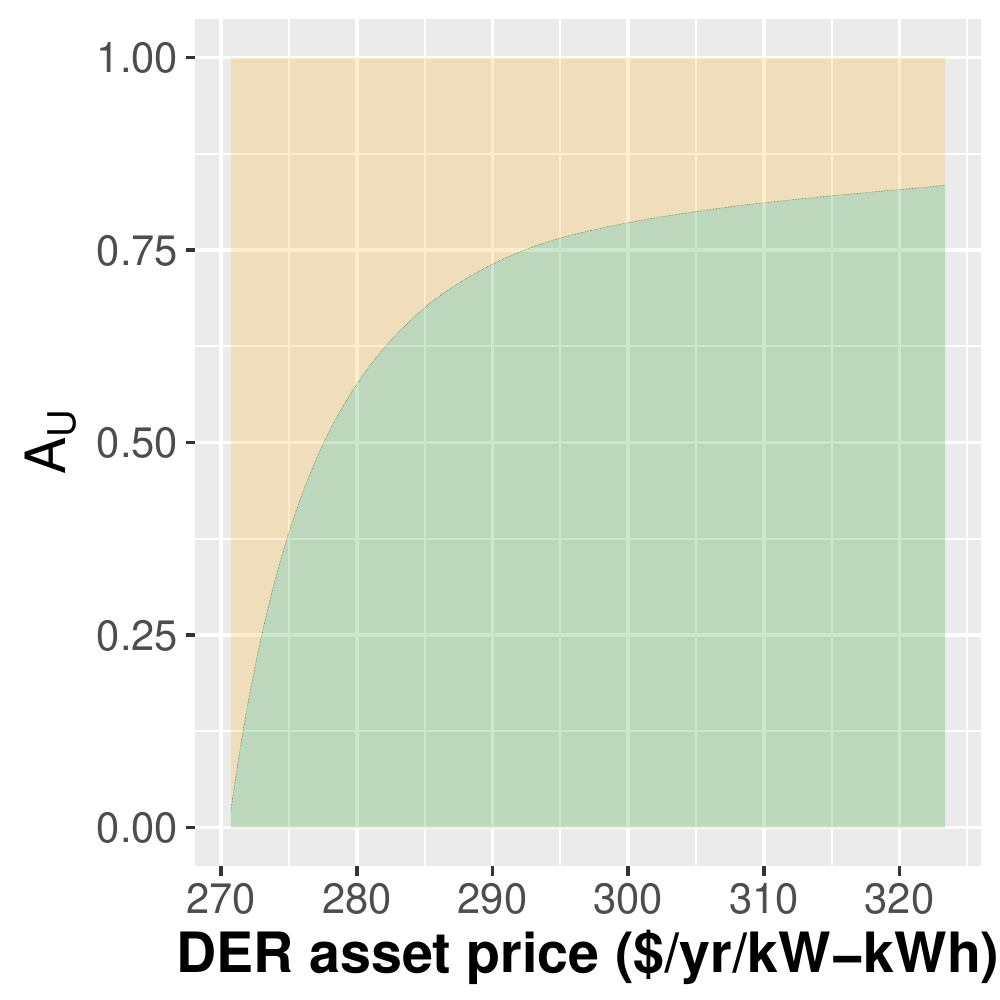}
\caption{}
\label{fig:fixed_price_regimes}
\end{subfigure}
\caption{(a) The green curve is the equipment vendor's increase in revenue $\Delta R_V(p) $at a given asset price due to the existence of the peer-to-peer market. The orange curve is the utility's loss of billed revenue $\Delta R_U(p)$. The absolute difference between the two is about the same across the the range of DER asset prices considered in this analysis. (b) In the orange region, the utility's profit rate ratio $A_U$ is high enough that its profit loss is higher than the vendor's profit gain, so the utility will outspend the vendor and block the emergence of the peer-to-peer market. In the green region, the utility's lost profits are smaller than the vendor's profit gains, so the vendor will outspend the utility and ensure the existence of the peer-to-peer rental market.}
\label{fig:fixed_price_results}
\end{figure}

When the purchase price $p$ of the DER asset is above about \$270/yr/kW-kWh, the peer-to-peer market leads to a long run increase in adoption, so the vendor's revenue increase $\Delta R_V(p)$ is positive.
The utility's billed revenue decreases by $\Delta R_U(D(p),\tilde{D}(p))$.
The revenue changes for the two entities are shown in Figure \ref{fig:fixed_price_DR}.
At all prices, the utility's loss exceeds the vendor's gain.
Depending on their rates of profit, one or the other may have more resources at their disposal to enable or block the emergence of the P2P market.

Figure \ref{fig:fixed_price_regimes} illustrates the regime in which the peer-to-peer market will come into existence as a function of $A_U$ and $p$.
As the asset price decreases, the utility needs a smaller profit rate ratio in order to block the rental market because its billed revenue losses are much higher than the vendor's revenue gains.
The utility never needs to have a higher rate of profit than the vendor (i.e. $A_U>1$) in order to block the market, but at higher asset prices, it needs a profit rate of at least 80\% of the vendor's.

\subsection{Comparison to direct subsidies}

\begin{figure}
\centering
\includegraphics[width=0.6\textwidth]{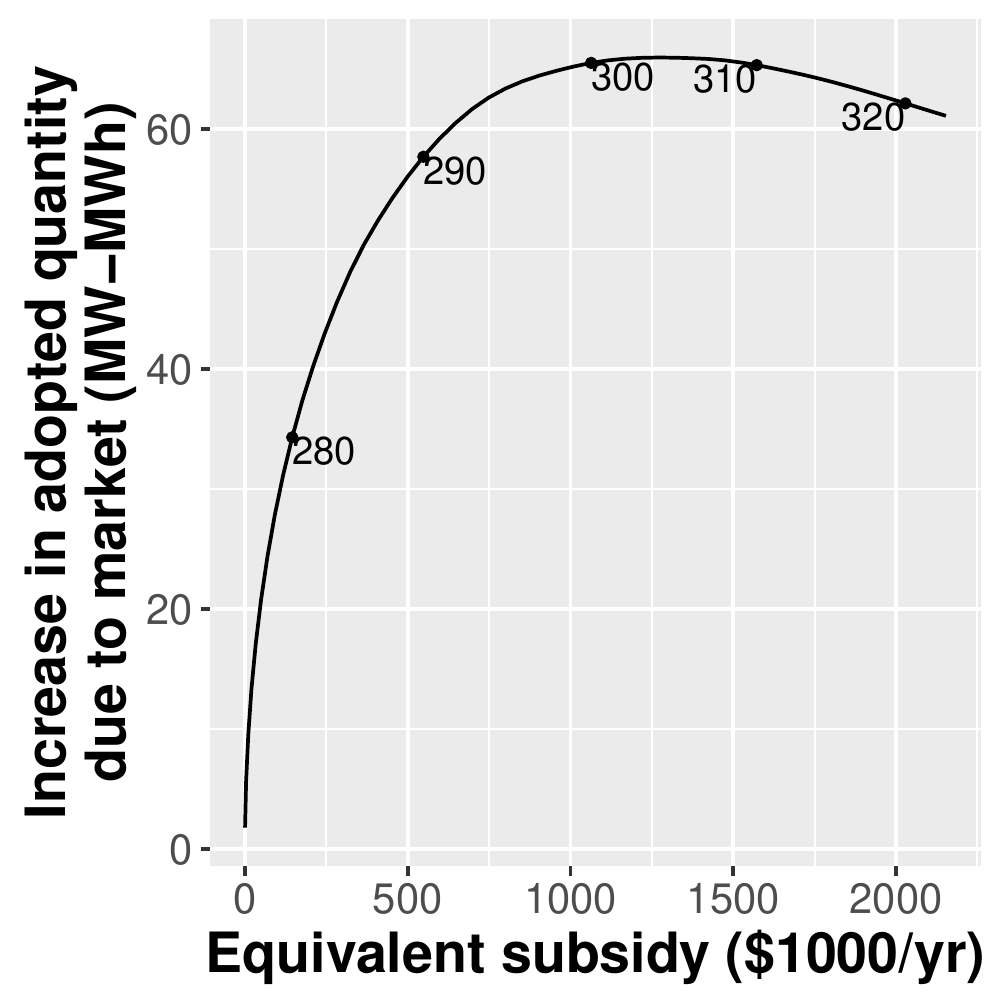}
\caption{The curve traces out the equivalent subsidy $S_E(p)$ that generates the same increase in adoption $\Delta Q(p)$ as the peer-to-peer market for different asset purchase price points $p$.
The asset purchase price varies along the curve, with particular values denoted by the labeled points.
As an example, suppose the asset purchase price is \$290/kW-kWh.
The peer-to-peer market generates an increase in total adopted quantity of around 57 MW-MWh, read off on the y-axis.
In the absence of the peer-to-peer market, the direct subsidy required for that same increase in adoption would be a little over \$500,000/yr, read off on the x-axis.}
\label{fig:equiv_subsidy}
\end{figure}

We consider the trade-off facing a policymaker interested in increasing DER adoption through either enabling the peer-to-peer market or through direct subsidies.
Figure \ref{fig:equiv_subsidy} gives a sense of the increase in DER adoption that the peer-to-peer market could facilitate for different purchase prices, as well as the equivalent direct subsidies required to get the same increase in adoption without the peer-to-peer market.
The peer-to-peer market leads to the greatest increase in adopted quantity, about 66 MW-MWh, when the purchase price is about \$305/yr/kW-kWh.
If, in lieu of enabling the peer-to-peer market, the policymaker wanted to generate the same increase through direct subsidies, the annualized cost would be about \$1.25 million per year.

\subsection{Geographical differences}

\begin{figure}
\centering
\begin{subfigure}{.49\textwidth}
\centering
\includegraphics[width=\linewidth]{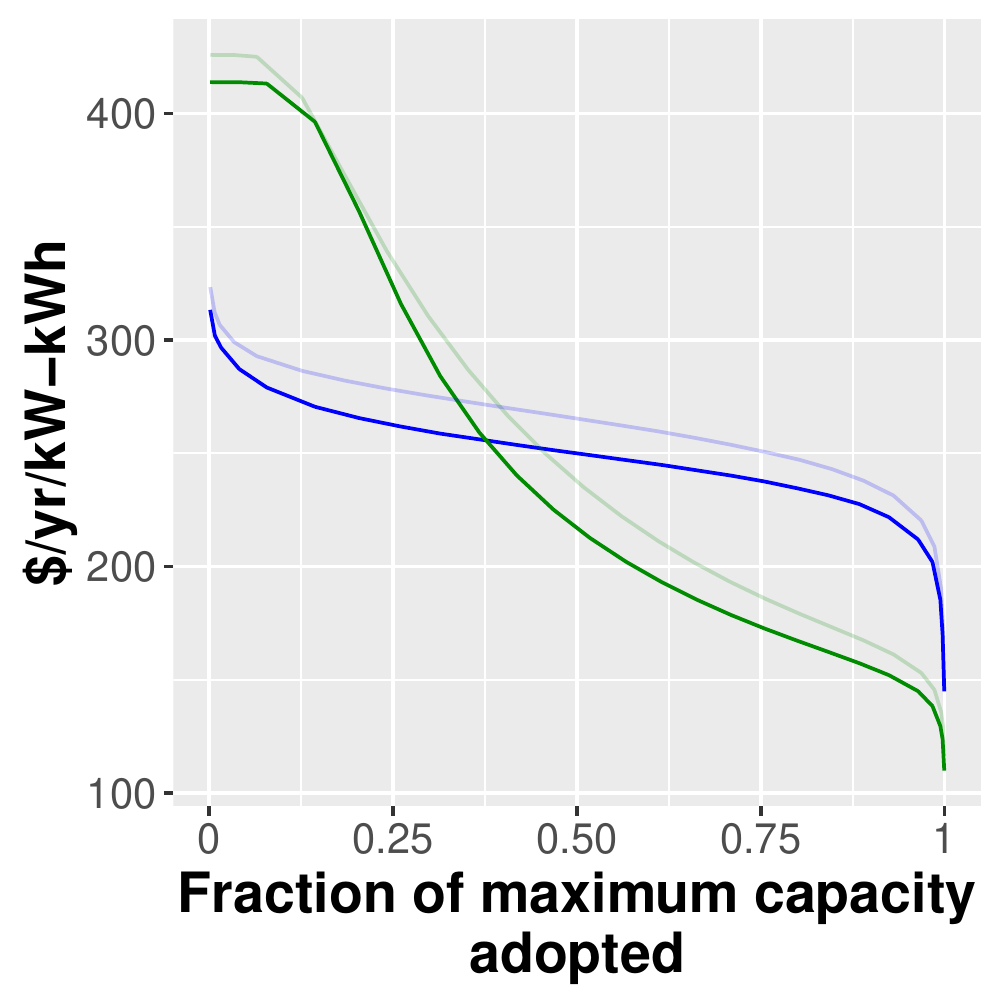}
\caption{}
\label{fig:Fresno_v_NEBay_adopted_frxn}
\end{subfigure}%
\begin{subfigure}{.49\textwidth}
\centering
\includegraphics[width=\linewidth]{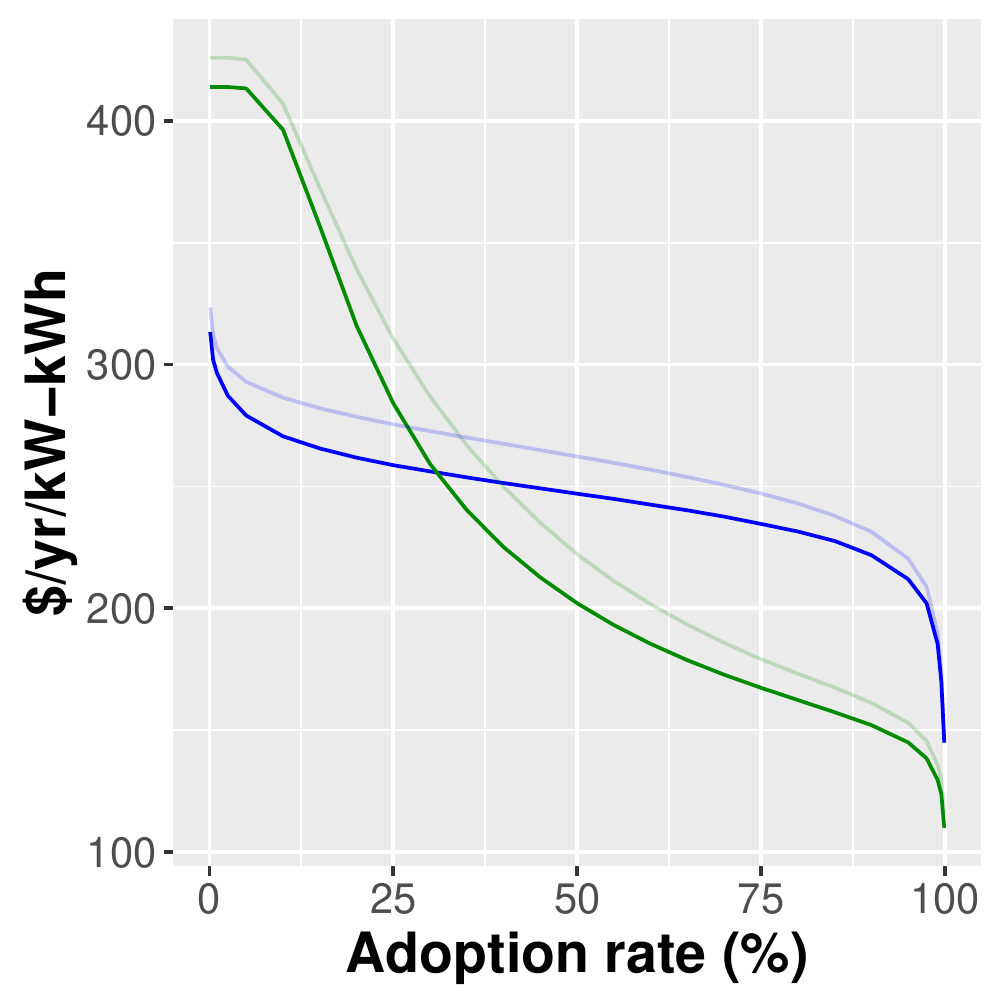}
\caption{}
\label{fig:Fresno_v_NEBay_adoption_pct}
\end{subfigure}
\caption{The darker lines are the demand curves in the Pittsburg area, and the lighter lines are those for the Fresno area. The blue curves are the demand in the absence of the peer-to-peer market, and the green curves are the demand with the market. The two areas have different numbers of adopters, and a different relationship between adoption rate and adopted quantity. Therefore, the most consistent comparison is given in (a), which  shows the demand as a function of the fraction of the maximum capacity adopted. The maximum capacity is defined as the total net-zero system size for all households in the region, which is 226 MW-MWh for the Fresno region and 138 MW-MWh for the Pittsburg region. The demand curves as a function of the adoption rate are given in (b).}
\label{fig:Fresno_v_NEBay}
\end{figure}

For the same November 2011 to October 2012 time frame, we have smart meter data for a group of 34,000 households in the Pittsburg, California area that experience very similar solar irradiance.\footnote{The slope coefficient for an ordinary least squares regression between the solar irradiance for any two zip codes for this set of households around Pittsburg ranges from  0.97 to 1.00, and the maximum mean absolute difference is 19 Wh/$\textrm{m}^2$. For the LMPs, the slope coefficient ranges from 0.81 to 1.11, and the maximum mean absolute difference is 0.2 \textcent/kWh. We generate a single weighted solar irradiance and LMP just as we did for Fresno.}
Figure \ref{fig:Fresno_v_NEBay} shows that the demand curves for Pittsburg are lower than those for Fresno, both with and without the peer-to-peer market.
Pittsburg receives 97\% of the total annual solar irradiance that Fresno does, and the weighted LMPs used in our study are practically identical between the two.
The median energy consumption of the households in Pittsburg is 76\% of that of the households in Fresno.
Recall that the DER system size for a household scales with its consumption, so Pittsburg's lower electrical energy consumption alone does not account for the lower demand.
The driver for the difference in demand between the two regions is the difference in the household's fraction of consumption that takes place during the peak hours of the TOU rate.
The Fresno households consume more of their energy during peak hours than those in Pittsburg, giving the same unit of DER asset capacity more scope to reduce bills, and therefore higher value, in Fresno.

\section{Discussion}

The peer-to-peer market enables many households to share in the benefits of distributed energy technologies, even when only a small fraction choose to own the DER asset themselves.
Over a wide range of adoption rates, there is robust participation in the market by owner and non-owner households.
In addition, up to about 15\% adoption, the peer-to-peer market achieves about as much total household bill savings as a centralized coordinator would.
The peer-to-peer market has the advantage of being more interactive and of having a built-in allocation mechanism for the bill savings.

The market clears very locally from about 10\% adoption up to very high adoption rates.
That means that rentals can often be matched to take place between neighbors, which bolsters the appeal of the peer-to-peer market as a way for communities of people to share in the benefits of technology.
It also means that the distribution grid operator does not have to be concerned about the rental market requiring large transfers of energy across large distances, or virtual assignments of storage device from one location to another far away.
This localness should be associated with lower cost impacts to the distribution grid, due to lower losses and energy back-flows over shorter distances.

The peer-to-peer market can facilitate an increase in adoption.
As an example, in our study, if the annualized purchase price of the DER asset hits \$290/yr/kW-kWh, about 5\% of Fresno households would have an incentive to purchase the asset for their own use.
The peer-to-peer market would make it profitable for more households to adopt the asset, with the rent-own equilibrium at about 30\% adoption.
Thus, policymakers seeking to increase DER adoption should consider enabling the peer-to-peer rental market.

Enabling the peer-to-peer market may entail costs of its own, which can be compared to the alternative of direct subsidies.
When comparing the two alternatives, we will assume that they involve the same increase in adopted DER quantity by the same households.
In other words, the additional rooftop PV and storage ends up in the same place in both cases.
Thus, the energy generated by the rooftop PV systems is produced in the same locations.
The storage will be operated differently, however, in the two cases.
This will lead to differences between the two alternatives in terms of inflows and outflows between a household and the distribution grid.

In our study, households get a very low price (i.e., the wholesale rate) for selling electricity back to the utility.
In the absence of the peer-to-peer market, owner households are thus strongly incentivized to consume as much of their generated energy as possible by using the storage device to store extra energy generated during the daytime for consumption at a later hour.
The household wants to minimize outflows.
By contrast, when an owner household rents part of its rooftop PV capacity to another on the peer-to-peer market, any physical energy transfer between the two happens over the wires of the distribution grid.
It involves an outflow from the household with the PV system because the rental agreement does not change the panel's physical location.
Distribution grid operators are concerned in particular with back-flows and voltage variations that may be driven by outflows from a household, which may result in increased upgrade or maintenance costs.

Furthermore, there may be purely virtual transfers of capacity.
It is possible that a household may rent out its entire PV capacity on the market.
Thus, even though it will certainly consume some of that energy locally, its generated energy is credited for billing purposes to another household's account.
Similarly, a household that rents storage capacity on the peer-to-peer market will in reality be controlling a storage device located elsewhere --- so the charging and discharging of the device will happen in one location but be credited elsewhere.
\footnote{Given that the storage device capacity is fungible on the rental market, the peer-to-peer market operator and the distribution grid operator could collaborate to optimize which storage devices respond to charging and discharging commands from renters. This could lead to reduced costs for the distribution grid.}

In our model, these peer-to-peer arrangements assume free use of the distribution grid and no losses or other physical constraints.
The distribution grid operator would have to deal with both realities.
The detailed rate sheet for the retail time use rate used in this study assigns about 45\% of the total billed revenue to distribution costs.
For the Fresno households in this study, that comes out to \$45 million of their total electricity bills prior to any DER adoption.
If the additional distribution grid costs related to enabling the peer-to-peer market exceed a few percent of original total distribution costs, then direct subsidies may be a cheaper way for a policymaker to achieve the same increase in adoption.

It should be noted, however, that the peer-to-peer market may help reduce distribution grid costs.
The rental market encourages homes with solar panels to export surplus energy instead of saving it for later.
Their neighbors can consume that energy, so the utility doesn't need to provide as much for the neighbors.
Distributing a unit of energy from one home to its neighbor is less resource-intensive than transporting that same unit of energy from a distant large scale generator, so the distribution costs for that unit of energy should be lower.
Of course, the generating home will have less energy stored for later, so there will be distribution costs for supplying the energy to make up that difference later in the day.
However, if the neighbor-to-neighbor sharing occurs at times when the distribution grid is relatively stressed, the overall cost of distribution may go down.
Thus, there is a reasonable case to be made that the rental market will cause distribution costs to come down because it incentivizes generating homes to supply their neighbors' needs.
This incentive structure is related to the pricing policy faced by the households.
The policy in our model offers households a relatively low price --- the wholesale rate --- for energy they sell back to the utility.
If the utility instead offered households a higher price for surplus energy, close to the purchase price, they wouldn't have much incentive to store it for later, so they would end up exporting it out to the grid and supplying their neighbor's needs in that way.
The rental market would have much less scope.

The operator of the peer-to-peer market platform needs cooperation from retail electric utilities for virtually reassigning the capacity and the actions of DERs to reflect the rental contracts.
The retail utility may hesitate to cooperate fearing a decline in revenue and an increase in cost and complexity of managing the distribution grid.
Aggregators that provide coordination services may also oppose the creation of peer-to-peer markets because they compete directly with the value stream that coordination can generate.
Equipment vendors would support the peer-to-peer market as a means to increase sales.
Policymakers will need to take into account these competing agendas and their underlying economic motivations, as well as the interests of residential consumers, when evaluating peer-to-peer rental mechanisms.

We suggest two extensions of this study.
The first would be an evaluation of the dynamics of the peer-to-peer market and how they would impact the static equilibria presented here.
Early levels of adoption would alter electricity prices, changing the incentives for later households considering adoption.
The retail utility may behave strategically, altering its rate structure in response to early adoption in a way that heads off later adoption.
In addition, households may require some minimum threshold surplus before they choose to participate in the market.
This friction may alter the character of the equilibria by excluding many renters and owners.

A second extension would be to enrich the set of motivations for households.
We limited our study to bill savings, but the desire to act more environmentally consciously may bolster rental demand.
On the other hand, it may also stifle rental supply, as owners may prioritize offsetting their own consumption with their net-zero systems over making them available to their neighbors.
A model that comprehensively accounted for household utility from bill savings, renewable consumption, and power security would be a valuable contribution.

\section*{Acknowledgements}

We thank Pacific Gas and Electric Company for providing the smart meter data used in this study.
We thank Dr. Ramesh Johari and Dr. Stefan Reichelstein for their valuable comments about the trade-offs involved in enabling the peer-to-peer market.

\bibliographystyle{IEEEtran}
\bibliography{bibtex/bib/DER_P2P}

\end{document}